**Global carbon stocks and potential emissions due to mangrove deforestation from 2000 to 2012**

Stuart Hamilton*

Department of Geography and Geosciences

Salisbury University

Salisbury

MD 21801, USA

sehamilton@salisbury.edu

+1 410-543-6460

Daniel A. Friess

Department of Geography

National University of Singapore

Singapore 117570

dan.friess@nus.edu.sg

+65 6516 1419

**Mangrove forests store high densities of organic carbon compared to other forested ecosystems. High carbon storage coupled with high rates of deforestation means that mangroves contribute substantially to carbon emissions. Thus, mangroves are candidates for inclusion in Intended Nationally Determined Contributions (INDCs) to the UNFCC Payments for Ecosystem Services (PES) program. This study quantifies two datasets required for INDCs and PES reporting. These are annual mangrove carbon stocks from 2000 to 2012 at the global, national, and sub-national levels and global carbon emissions resulting from deforestation. Mangroves stored 4.19 Pg of carbon in 2012, with Indonesia, Brazil, Malaysia, and Papua New Guinea accounting for greater than 50% of this stock. 2.96 Pg of the global carbon stock is contained within the soil and 1.23 Pg in the living biomass. Two percent of global mangrove carbon was lost between 2000 and 2012, equivalent to a maximum potential of 316,996,250 t of $CO_2$ emissions.**

Forestry, agriculture, and other land use changes account for almost 25% (up to 12 Pg $CO_2$-e $yr^{-1}$) of anthropogenic greenhouse gas emissions, due to factors such as deforestation, forest degradation and biomass burning [1]. The deforestation of tropical coastal wetlands such as mangrove forests contributes disproportionately to anthropogenic greenhouse gas emissions, as they mangrove forests can hold up to four times as much organic carbon per unit area when compared to other terrestrial forested ecosystems [2] and are undergoing deforestation across the tropics [3, 4]. Recent estimates have put global mangrove deforestation rates at up to 0.39% per year since 2000 [4], driven primarily by large-scale agricultural and aquacultural commodity production, [3, 5, 6, 7] coastal development [3, 5, 6, 7], and sea level rise [8]. High carbon densities per unit area coupled with high deforestation rates mean that globally mangrove deforestation may be contributing as much as 0.21 Pg $CO_2$-e $yr^{-1}$ or 0.45 Pg $CO_2$-e $yr^{-1}$ to the atmosphere [2, 9]. Mangrove deforestation is so high in particular countries such as Indonesia that halting deforestation has been estimated to reduce its national land use sector emissions by between 10% and 31% [10]. As a result, carbon stored in coastal wetlands such as mangroves has recently been placed on the international policy agenda through the United Nations Framework Convention on Climate Change (UNFCCC) Paris Agreement in 2015 [11]. Due to this inclusion, emissions from wetlands are now explicitly considered in national greenhouse gas emissions reporting through the Intergovernmental Panel on Climate Change (IPCC)'s wetland supplement to the Guidelines for National Greenhouse Gas Inventories [12].

The Paris Agreement also provides new opportunities for mangrove conservation, as it promoted novel funding avenues for the financing of forest protection. Several conservation mechanisms have recently been established or proposed that utilize vegetated carbon stocks as a financial incentive to reduce deforestation, under the broad umbrella of PES. PES is broadly defined as a set of "voluntary transactions between service users and service providers that are conditional on agreed rules of natural resource management" [13 p.8]. For example, PES schemes such as Reducing Emissions from Deforestation and Degradation (REDD+) incentivize conservation through 'avoided deforestation,' with a service buyer paying a service provider to store carbon that would otherwise be emitted due to land cover change. Payments for avoided deforestation are increasingly advanced in terrestrial forest conservation, and such an approach is rapidly gaining traction in mangrove research and policymaking under the term "blue carbon" [14, 15]. Blue carbon is quickly gaining international prominence as a conservation tool through groups such as the International Blue Carbon Initiative and is the focus of several bilateral government frameworks, such as the International Blue Carbon Partnership between Australia and Indonesia. Case studies have shown that the financial benefits accrued from the sale of blue carbon credits could potentially outweigh financial returns from alternative land uses at the local scale [16], and thus provide an economically viable alternative to some proximate drivers of mangrove deforestation and degradation.

The calculation of both the emissions from land cover change in national greenhouse gas inventories, and the calculation of ecosystem service loss for PES interventions require robust information on standing vegetated carbon stocks, and emissions due to land cover change through time. For example, most definitions of PES need some form of conditionality [13], which sets rules and standards that must be met by the service provider for payment to be made. Thus, financial transactions under PES require robust information on variables such as carbon storage and rates of habitat loss, to allow the accurate quantification of carbon credits and carbon saved through avoided deforestation. We particularly need to know baselines of deforestation and carbon storage at varying spatiotemporal scales, from the site to the national level. However, we currently lack robust baselines of mangrove deforestation in many countries across the tropics [17]. We also require robust estimates of mangrove carbon stocks and emissions due to deforestation at multiple scales. Our lack of information on these parameters, at the local, national and regional levels hampers the efforts of decision-makers to calculate emissions and suitable reduction mechanisms or set adequate baselines of loss, from which to assess the effectiveness of a PES intervention [18].

This study reports global, national, and sub-national mangrove carbon stocks for the year 2012 and estimates the global carbon stock losses and potential $CO_2$ emissions resulting from mangrove area change between 2000 and 2012. In addition to reporting global mangrove carbon stocks, it delineates the amount of global mangrove carbon in the aboveground living pool, the belowground living pool, and in the mangrove soil; key carbon pools that must be delineated and quantified for national emissions reporting [19]. It reports mangrove stocks at national and sub-national scales and for the first time makes available to researchers the geospatial data required to track the change in mangrove carbon stocks at the transnational, national, and sub-national scales.

This study provides robust baseline information on carbon stocks for use in national emissions reporting and PES schemes at high spatiotemporal resolutions. This study advances previous efforts [20, 21] that estimated carbon emissions from mangroves, by employing the most recent and state-of-the-art datasets available on deforestation and carbon stocks for all carbon pools at the global scale to provide high-resolution, global, robust, and transparent calculations of potential carbon emissions from mangroves. For example, when comparing our estimate to the most comprehensive current global mangrove carbon estimates [20, 21] we utilize: (i) 13-years of data as opposed to 1-year [20, 21] (ii) five differing biomass models that include latitude models, environmental parameter models, and field derived models as opposed to using only one model that is either latitudinal [21] or field based and localized [20] (ii) a spatial resolution of approximately 0.0009 $km^2$ at the equator and opposed to using a grid of 81 $km^2$ [21] which allows for sub-national mangrove carbon estimate, (iii) more recent mangrove area data from 2012 as opposed to mangrove are data from 1997 to 1999 [21], (iv) one hundred potential mangrove cover measures at each minimum mapping unit as opposed to presence and absence data at the pixel level [21], and (v) a fully open data distribution system for all global data at the pixel level as opposed to not having the data available [20, 21] for other researchers to replicate, validate, or utilize.

We find that the global mangrove carbon stock in 2012, assuming a conservative but standard 1 m average soil depth, as per the IPCC [12] and other global studies [2, 21, 22], was estimated to be 4.19 ± 0.62 Pg (CI 95%). Of this 4.19 Pg, 2.96 ± 0.53 Pg of the global mangrove carbon stock is contained within the soil, and approximately 1.23 ± 0.06 Pg is in the standing and living mangrove biomass. Of this 1.23 Pg, approximately 0.41 ± 0.02 Pg is belowground biomass in the root system and approximately 0.82 ± 0.04 Pg is in the aboveground living biomass (Figure 1). This equates to approximately 70.65% of global mangrove carbon being contained in mangrove soils, 9.78% in belowground biomass and 19.57% in aboveground biomass.

**Figure 1. Global distribution of mangrove ecosystem carbon stocks per pool for the year 2012.**

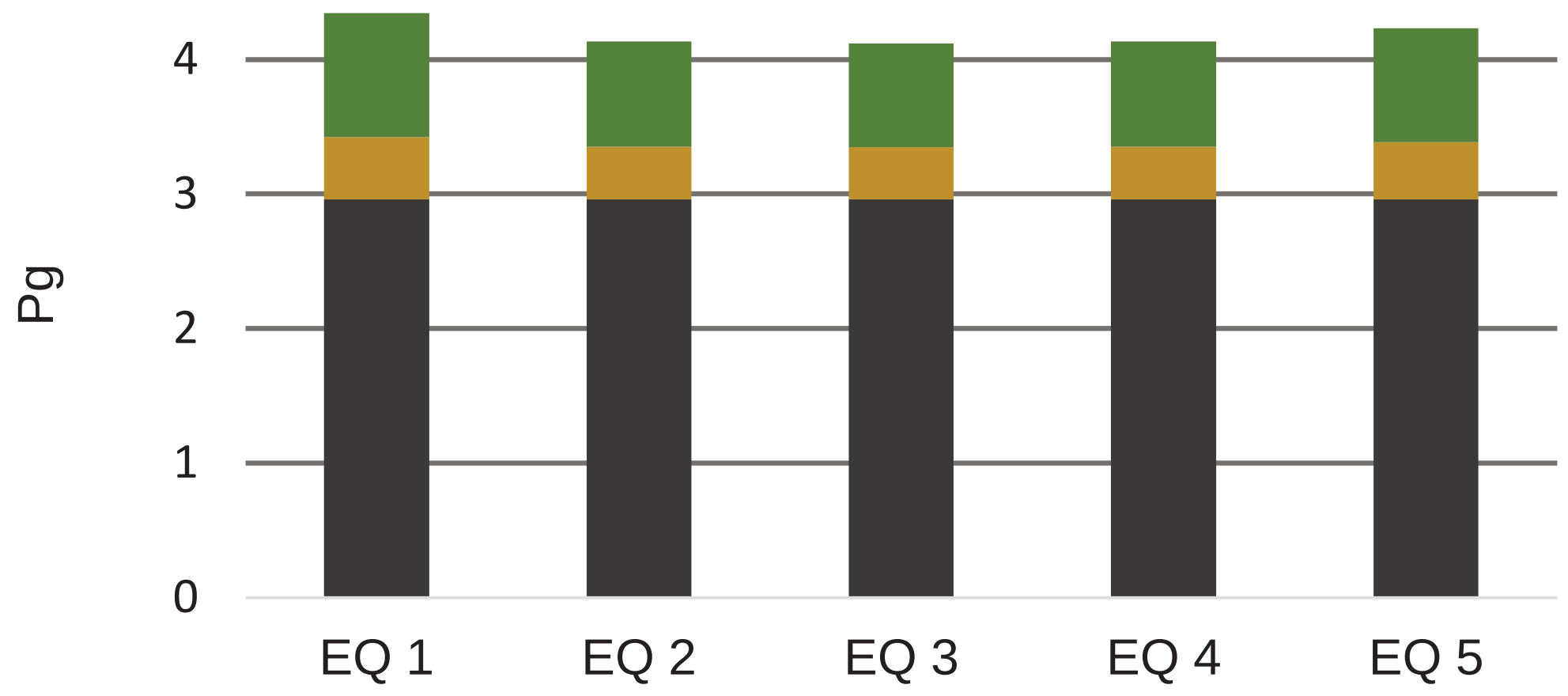

Global distribution of mangrove ecosystem carbon stocks per pool for the year 2012. The lower gray portion of each bar represents soil carbon. The central brown portion of each bar represents belowground living carbon in the root system. The upper green portion of each bar represents aboveground living carbon in the tree.

Indonesia, Brazil, Malaysia, and Papua New Guinea contain more than 50% of the world's mangrove carbon stock, with Indonesia alone accounting for more than 30% of the entire world's mangrove carbon stock (Table 1). The top 10 mangrove holding countries contain just under 70% of the world's mangrove carbon stocks and the top 25 just over 90%. Interestingly, countries with large mangrove areas do not always have equivalently large mangrove stocks. For example, Bangladesh ranks three places lower globally when ranked by mangrove carbon stocks than if it were ranked by actual mangrove area. Conversely, Gabon is ranked two spots higher than if it were ranked solely based on mangrove area (Table 1).

**Table 1. National estimates of mangrove carbon holdings, 2012.**

| Country Name | Mangrove Area (2012) km$^2$ | Mangrove Area Rank | Tonnes of Carbon | | | Percent of Global Total | C Rank | Change |
|---|---|---|---|---|---|---|---|---|
| Indonesia | 23,324.29 | 1 | 1,275,115,175 | ± | 19,597,086 | 30.41 | 1 | 0 |
| Brazil | 7,674.94 | 2 | 389,760,564 | ± | 9,556,539 | 9.30 | 2 | 0 |
| Malaysia | 4,725.84 | 3 | 258,882,085 | ± | 4,002,528 | 6.17 | 3 | 0 |
| Papua New Guinea | 4,172.29 | 4 | 223,096,105 | ± | 3,836,601 | 5.32 | 4 | 0 |
| Australia | 3,316.21 | 5 | 152,539,573 | ± | 2,104,454 | 3.64 | 5 | 0 |
| Mexico | 2,991.83 | 6 | 149,261,592 | ± | 1,203,826 | 3.56 | 6 | 0 |
| Nigeria | 2,653.99 | 7 | 127,914,456 | ± | 2,559,377 | 3.05 | 7 | 0 |
| Myanmar | 2,557.45 | 8 | 118,883,668 | ± | 1,409,261 | 2.84 | 8 | 0 |
| Venezuela | 2,403.83 | 9 | 112,537,865 | ± | 1,851,142 | 2.68 | 9 | 0 |
| Philippines | 2,064.24 | 10 | 104,470,697 | ± | 1,341,367 | 2.49 | 10 | 0 |
| Thailand | 1,886.33 | 11 | 91,793,396 | ± | 1,414,284 | 2.19 | 11 | 0 |
| Colombia | 1,671.86 | 13 | 84,108,157 | ± | 1,831,402 | 2.01 | 12 | 1 |
| Cuba | 1,633.46 | 14 | 81,223,503 | ± | 651,189 | 1.94 | 13 | 1 |
| USA | 1,568.60 | 15 | 75,453,694 | ± | 622,606 | 1.80 | 14 | 1 |
| Bangladesh | 1,772.98 | 12 | 74,049,402 | ± | 653,854 | 1.77 | 15 | -3 |
| Panama | 1,323.94 | 16 | 72,923,978 | ± | 1,222,387 | 1.74 | 16 | 0 |
| Gabon | 1,082.11 | 19 | 58,592,889 | ± | 1,979,216 | 1.40 | 17 | 2 |
| Mozambique | 1,223.67 | 17 | 55,803,315 | ± | 723,403 | 1.33 | 18 | -1 |
| Ecuador | 935.74 | 20 | 55,566,461 | ± | 1,660,042 | 1.33 | 19 | 1 |
| Cameroon | 1,112.76 | 18 | 53,980,215 | ± | 1,138,012 | 1.29 | 20 | -2 |

Supplemental Table 1 extends Table 1 to all 105-mangrove holding level one administrative units globally. The plus or minus data only accounts for living carbon.

Although national estimates of mangrove carbon stocks are important, it is at the sub-national level that these data likely have the most utility. This is important as slightly over one-third of the global mangrove carbon stocks are contained within only ten level one administrative units (Table 2). Indeed, greater than 50% of the world's mangrove carbon stocks are located within only 21 administrative level units (Supplemental Table 2). Level one organizational units are typically one level below the nation state. Within the USA, the level one administrative unit is the state or governed territory, in Indonesia, they are the 34 provinces, and in Australia, it would be the states and any administered territories. Five of the top ten mangrove carbon holding level-one administrations, including the first and second, are in Indonesia. The top ten administrative level one units contain slightly over one-third of the entire global mangrove stock. As opposed to national estimates, computational limitations make it difficult for standard deviations, and hence confidence intervals, to be generated for all possible sub-national boundaries and are therefore not included in sub-national estimates. Although individual researchers can produce confidence intervals from the provided database for their sub-national areas of interest.

**Table 2. Administrative level one unit estimates of mangrove carbon holdings, 2012.**

| Country | Level One Administrative Unit | C t | % of Global Total |
|---|---|---|---|
| Indonesia | Papua | 328,816,690 | 7.77% |
| Indonesia | Papua Barat | 237,459,220 | 5.61% |
| Brazil | Maranhão | 155,013,142 | 3.66% |
| Malaysia | Sabah | 137,359,199 | 3.25% |
| Papua New Guinea | Gulf | 122,124,709 | 2.89% |
| Brazil | Pará | 106,739,631 | 2.52% |
| Indonesia | Kalimantan Timur | 95,815,540 | 2.26% |
| Indonesia | Maluku | 92,862,422 | 2.19% |
| Myanmar | Tanintharyi | 87,519,738 | 2.07% |
| Indonesia | Sumatera Selatan | 81,738,975 | 1.93% |

Sub-national estimates use EQ 5 from the methods section and the mid-level mangrove AGB to BGB conversion ratio. Supplemental Table 2 extends Table 2 to all 752-mangrove holding level one administrative units globally.

Many mangrove forests are managed at highly granular levels beyond commonly mapped administrative units. For example, mangrove holding nations such as Indonesia, India, Thailand and the Philippines have highly successful community-based mangrove management programs, and other countries such as countries such Cambodia, Sri Lanka, Bangladesh, Iran, Honduras, Ecuador, Brazil, and Panama have committed themselves to community-based mangrove management programs [23]. The data generated in this study can likely provide useful carbon stock estimates even at management-relevant scales.

Globally, mangrove carbon stocks have decreased by a maximum 86,375,000 ± 1,367,000 t C or slightly over 2.0% during this period (Figure 2). This assumes a total carbon loss when mangrove deforestation occurs. For example, if 500 $m^2$ of mangrove forest is lost then we assume total mangrove carbon losses for this 500 $m^2$. Carbon stock loss rates are highly consistent across the 13-year analysis period, averaging 0.17% per year. Although the annual rate of decline from 2000 to 2012 is consistent globally, loss at the country level differs substantially. For example, Indonesia alone is responsible for almost 41,946,838 t, or 48.56%, of the entire global mangrove forest carbon stock loss during this period. Also, Myanmar has a loss rate of 7.99%, a four-fold increase over the global loss rate for the period 2000 to 2012, so contributes a disproportionate volume of emissions relative to its total mangrove extent. Indeed, Southeast Asia is identified a hotspot of global mangrove carbon stock losses between 2000 and 2012.

**Figure 2. Global mangrove carbon losses, 2000 – 2012.**

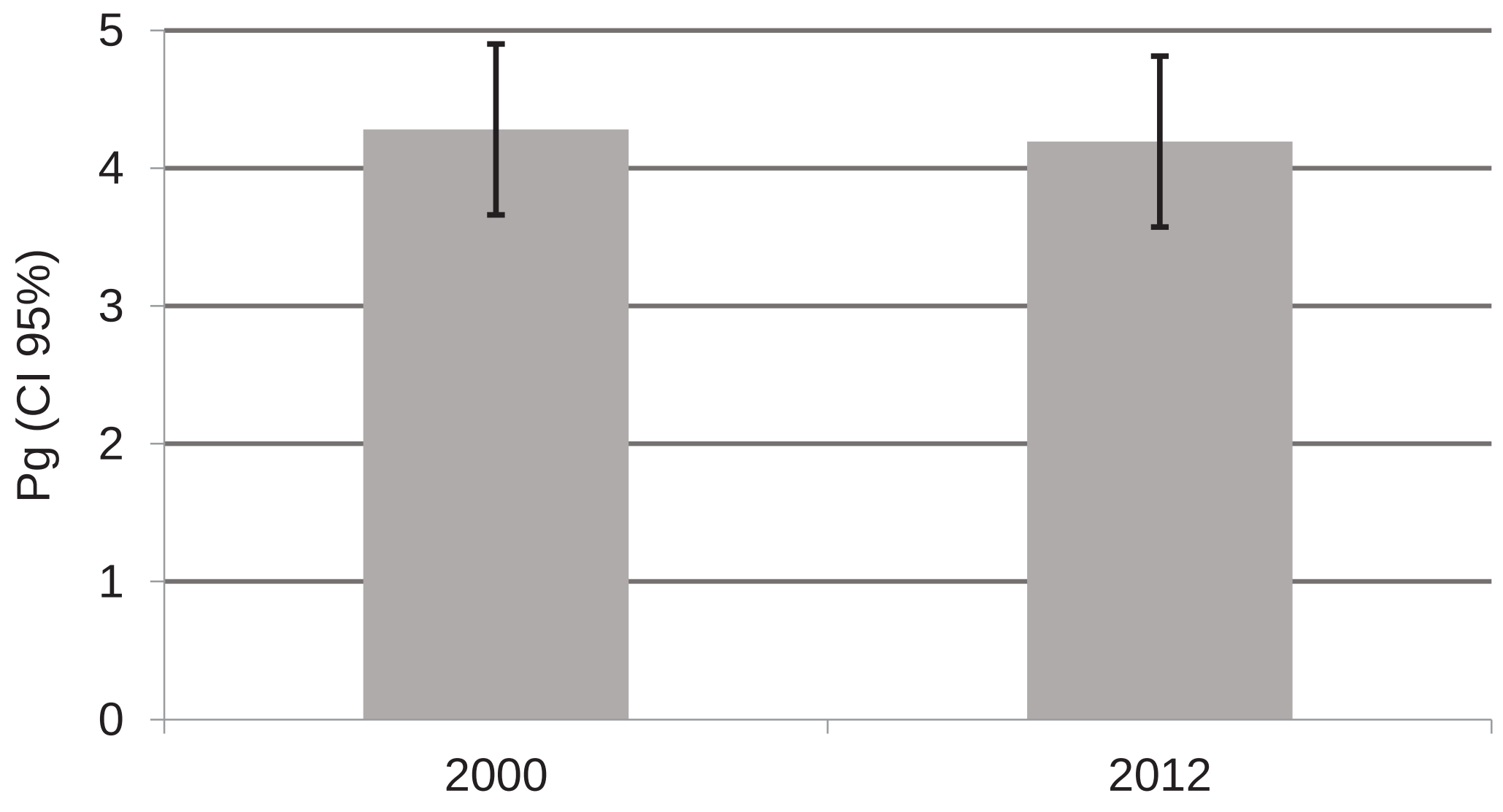

Using the assumption that between 25% and 100% the total amount of mangrove clearing is converted into $CO_2$ emissions[9], the global carbon losses reported could have resulted in a potential maximum increase in mangrove clearing-induced $CO_2$ emissions of between 79,249,063 and 316,996,250 t annually.

This higher number is equivalent to the $CO_2$ emissions of approximately 67.5 million US passenger vehicles annually [24]. In addition to carbon losses during the analysis period, these data can also be used to estimate the amount of $CO_2$ that these deforested mangroves are no longer removing from the atmosphere. For example, using an average annual manual carbon sequestration rate of 163 (+40, -41) g OC $m^{-2}$ $yr^{-1}$ (95% CI)[25] and 210 g $CO_2$ $m^{-2}$ $yr^{-1}$ [26], and recent mangrove loss estimates[4]. We estimate that cleared mangrove forests between 2000 and 2012 have not sequestered an additional 3,487,874 t to 4,493,580 t of carbon over the analysis period had deforestation not taken place. Such sequestration losses occur in perpetuity.

The global mangrove carbon estimate produced using this methodology lowers the current consensus global mangrove carbon stock estimate. For example, Siikamäki, Sanchirico [21] estimated the year 2000 mangrove carbon stocks at 6.5 Pg of carbon and Jardine and Siikamäki [27] estimated the year 2000 mangrove soil carbon alone to be 5.00 ± 0.94 Pg. This compares to 4.28 ± 0.62 Pg of total carbon stocks in 2000 estimated by this current study (Figure 2). The lower estimate produced is likely due to three interacting reasons related to the earlier use of presence and absence tree cover data, the spatial resolution of the differing analysis, and the equations utilized. The supplemental discussion expands upon the potential reasons for these differences.

In addition to the global-scale overview of mangrove carbon stocks, robust and verifiable national and regional trends of mangrove forest carbon stocks can be estimated annually for a 13-year period. Information at national scales is required for accurate national emissions reporting [19, 28]. Nations can report activity data (the magnitude of human activity resulting in emissions, as reported in this study) across three tiers of increasing methodological complexity and reliability, with countries recommended to pursue estimates at the highest tier possible depending on data availability [29]. Most countries calculate activity data and emissions factors using default numbers provided by the IPCC (Tier 1), though studies such as that presented here allow us to create more precise and reliable estimates of carbon emissions by introducing spatial variation in both carbon stocks and land cover change (Tier 2). Indeed, these data are likely even applicable to Tier 3 reporting standards, which require repeated observation over time driven by high-resolution subnational granular spatial data that provides measures such as biomass and soil dynamics [30]. At Tier 3, higher order methods are used, including models and inventory measurement systems tailored to address national circumstances, repeated over time, and driven by high-resolution activity data and disaggregated at subnational to fine grid scales. These higher order methods provide estimates of greater certainty than lower tiers and have a closer link between biomass and soil dynamics.

Information on carbon stock trends is also of use at lower scales in the context of PES. Depending on the precise mechanism, PES is currently being pursued at both the national level, where ecosystem service payments are made to a national government (as often promoted by the UNFCCC in the context of REDD+), or at the site- or sub-national scales, where payments accrue to decision makers in specific jurisdictions [31]. At present, sub-national PES approaches may be more successful in the context of mangrove conservation, because the geographical position of mangroves in the contested intertidal zone provides many governance challenges at the national level in many tropical developing nations, especially in Southeast Asia [32]. This study allows for monitoring of PES (e.g., REDD+) and other conservation interventions programs at sub-national scales never available and can be used to generate necessary baseline surveys for such interventions. Indeed, this study protocol allows mangrove forest carbon levels to be estimated at regular intervals both post-intervention and pre-intervention during the PES process.

Both national emissions inventories and monitoring for PES compliance requires robust datasets that go beyond the 2000 to 2012 period presented. Moving forward, it is essential that research continues to produce robust, standardized data on carbon stocks and their losses and gains at scales that are relevant to policy makers. This requires extending the timeline of this present study backward to calculate longer historical baselines of emissions for inventories and the setting of PES baselines [33] and forward to test the efficacy of future conservation interventions [34] that protect mangrove carbon stocks.

Although considerable local variation occurs in mangrove carbon stocks losses year on year, at the global scale the losses are consistent confidence in near-term predictions. By 2017 we anticipate carbon global stocks to have decreased to 4.16 Pg ($y = -0.0074x + 4.2901$, $R^2$ 0.99, were x is the last two years of the year and is the global mangrove carbon stock). Suppl. Fig. 1 has the full global yearly data.

## Methods

*Global mangrove extent*

The mangrove carbon estimations generated begin with a remotely sensed global measure of mangrove cover in square meters at 1 arc-second spatial resolution; this is approximately 30 m at the equator. The mangrove forest cover estimation was taken from the Continuous Mangrove Forest Cover for the 21st Century (CGMFC-21) database that monitored mangrove canopy cover globally and annually from 2000 to 2012 [35]. Each of the 123,332,913, 1 arc-second pixels in CGMFC-21 contains a measurement of mangrove canopy cover in square meters with a minimum value of 0, indicating deforestation during the analysis period, and a maximum value of 955 $m^2$, indicating a pixel with total mangrove forest canopy cover in close approximation to the equator. Data omissions are shown to be less than 0.01% of the global mangrove area [35]. CGMFC-21 is a synthesis product that combines the remotely-sensed Global Forest Cover database [36], the remotely-sensed Mangrove Forests of the World (MFW) database [37], and the expert-compiled Terrestrial Ecosystems of the World database [38]. The CGMFC-21 GIS mangrove cover data are available to download from http://bit.ly/1lMJ9zj.

*Above- and below-ground biomass*

From the estimation of mangrove cover, above-ground biomass (AGB) was derived using a series of five latitudinal or bioclimatic linear equations present in the literature that relate AGB to a combination of latitude, geographic region, the mean temperature of warmest quarter, the average temperature of the coldest quarter, and precipitation of driest quarter (Table 3). Each equation was processed for all 123,332,913 pixels with mangrove forest cover in the CGMFC-21 database and adjusted to account for the square meter unit used in CGMFC-21. EQ 1 to EQ 3 are best fit linear models that relate field measures of AGB to latitude. EQ4 is a climatic model that relates AGB to numerous climatic variables. EQ 5 is a mean equation that averages both EQ 1 and EQ 3 to produce a value consistently within 1% of the average across all equations.

**Table 3. AGB equations used in this study, based on latitudinal and climatic variables.**

| NAME | SOURCE AND TYPE | EQUATION | r | P |
|---|---|---|---|---|
| EQ1 | Twilley et al. 1992 [39] [latitudinal] | AGB t pixel = (\|lat\| * -7.291 + 298.5) * CGMFC-21: AREA * 0.0001 | 0.75 | NR |
| | | | | |
| EQ2 | Hutchison et al. 2014 [40] [latitudinal] | AGB t pixel = (\|lat\| * -4.617 + 239.9) * CGMFC-21: AREA * 0.0001 | 0.37 | NR |
| | | | | |
| EQ3 | Saenger & Snedaker 1993 [41] [latitudinal] | AGB t pixel = (\|lat\| * -4.617 + 239.9) * CGMFC-21: AREA * 0.0001 | 0.69 | <0.0001 |
| | | | | |
| EQ4 | Hutchison et al. 2014 [40] [bioclimatic] | AGB t pixel (0.295BIO10 + 0.658BIO11 + 0.0234BIO16 + 0.195BIO17 −120.3) * 0.0001 * CGMFC-21: AREA | 0.53 | NR |
| | | | | |
| EQ5 | This study, average of EQ1 and EQ2 [39, 41] [latitudinal] | AGB t pixel = ((\|lat\| * -7.291 + 298.5) * CGMFC-21: AREA * 0.0001) + ((244.994 - (5.57 * \|lat\|)) * CGMFC-21: AREA * 0.0001) / 2 | 0.75<br>0.69 | NR<br><0.0001 |

AGB = above-ground biomass, t = metric tons, |lat| = absolute latitude, and CGMFC-21: AREA is the area of mangrove cover in m2 from the CGMFC-21 database. BIO10 = mean temperature of warmest quarter, BIO11 = mean temperature of the coldest quarter, BIO17 = precipitation of driest quarter.

EQ 1 is a linear latitudinal model [39], that forms the basis of many of the mangrove carbon estimates found in the academic literature at both the global e.g., [40, 42] and local scales [43, 44]. EQ 1 was developed using approximately thirty-five field measures of AGB globally.

EQ 2 is a second linear latitudinal model [40]. This model is a reparameterizing of EQ 1 using an expanded fifty-two global measures of mangrove AGB. After reparameterizing, EQ 1 results in an improved model fit with a decrease in AIC of 3.68 and an almost doubling of the variance explained by EQ 1 [40]. EQ 1 is likely applicable on a global scale although care should be taken in regions with few samples such as West Africa and the Pacific coastline of South and Central America.

EQ 3 is a third linear latitudinal model based on forty-three field measures of AGB[41]. Although it is used for some local estimates of mangrove carbon [43, 44] it does not appear to have the same prevalence in the academic literature as EQ 1. EQ3 is likely most suited for the heavily sampled Indo-West Pacific (IWP) region but should be treated with caution across Africa where no samples are taken, and across South America with only one sample.

EQ 4 differs substantially from EQs 1 to 3, as it is a climatic model based on three bioclimatic variables as opposed to calculating AGB purely as a function of latitude. It was developed by surveying the global bioclimatic database  to find relationships between the various bioclimatic variables and AGB [40]. The three bioclimatic variables that showed the strongest relationship to AGB were mean temperature of warmest quarter (BIO10), mean temperature of coldest quarter (BIO11) and precipitation of driest quarter (BIO17). These bioclimatic variables explained between 25.1% and 26.7% of the global variation in AGB [40]. The bioclimatic variables themselves were built from a global network of weather station data that were interpolated to the 30 arc-second level to provide a comprehensive global database of bioclimatic estimates [45]. EQ 4 is likely most suitable for global estimates as the underlying weather station data shows near global coverage [45], with some limited omissions in-and-around Kalimantan, North and West Sumatra, Papua (all Indonesia), the Horn of Africa, and portions of West Africa. Care should be taken in areas of sharp climatic transitions.

EQ 5 averages EQ 1 and EQ 2 at the individual pixel level. It has been shown to provide AGB estimations within 1% of field-measured, and allometric AGB estimates used to determine national scale mangrove carbon stocks in Ecuador [43]. This may be particularly useful as EQ 1, EQ 2, and EQ 3 may not be representative in this important mangrove–holding region. EQ 5 consistently produces results within 1% of the mean across all other Equations.

Mangrove BGB is typically calculated as an allometrically derived ratio of mangrove AGB [46, 47, 48]. We utilized a range of AGB : BGB ratios to estimate BGB at the pixel level. The low estimate comes from Hutchison, Manica [40] who found a global mean ratio of 0.39. The highest ratio of 0.61 was derived extracting an AGB to BGB ratio from 19 samples recorded by Komiyama, Ong [48]. The mid-range estimate is merely the mean of the high and the low, the mid-point ratio of 0.5, and aligns well with the ratio of 0.52 used in a review of conversion ratios used across the academic literature [44].

Conversion factors to convert whole-tree mangrove biomass to mangrove carbon exist in a narrow range of values between 0.45 and 0.50, based on the academic literature [39, 46, 49, 50]. For these calculations, we selected the mid value of 0.475 to represent the mangrove biomass to mangrove carbon conversion ratio.

*Soil carbon estimation*

To allow for complete ecosystem mangrove carbon stock assessments and to account for the fact that mangrove soils likely contain some of the highest carbon stocks of landcover type globally [49, 50, 51, 52, 53, 54], soil carbon estimations were included in the overall mangrove carbon calculations. Soil carbon values are estimated based on a predictive model of spatially explicit global mangrove soil carbon stocks [51]. This 5 arc-minute dataset calculated mangrove soil carbon stocks using a tree branching algorithm in a supervised machine learning environment based on the presence or absence of mangrove in circa 2000[37]. In a similar manner to EQ 4, soil carbon rates were estimated based on a relationship to bioclimatic variables in the bioclim database [45]. The soil model uses a Bag Decision Tree algorithm that generates the relative importance of each variable to the soil carbon estimate without establishing a linear relationship between the variables [55]. The relative importance of each bioclimatic variable, in addition to latitude and region, are reported in Table 4.

**Table 4. Soil carbon and bioclimatic variables relationship.**

| Variable Name | Variable Description | Relative Importance in Soil Carbon Estimation |
|---|---|---|
| BIO12 | Annual Precipitation | 20.92% |
| Latitude | Absolute Latitude | 19.67% |
| Region | Geographic Region | 17.32% |
| BIO15 | Seasonal Precipitation | 15.32% |
| BIO11 | Coldest Quarter Mean Temperature | 13.47% |
| BIO01 | Annual Mean Temperature | 13.31% |

The 5 arc-minute mangrove soil grid provides soil carbon values for each mapping unit in mg C per $cm^3$ [51]. The 1 arc-second grid in this study uses the 5 arc-minute mg C per $mm^2$ measure within which it is nested. The soil carbon measure is then adjusted to metric tonnes per 1 m depth. The multiplication by CGMFC-21: AREA in EQ 6 adjusts the data for mangrove coverage area in CGMFC-21 as opposed to mangrove presence or absence measure utilized [51].

(6)

$$\text{C t per 1 m}^3 = ((\text{Bag DT} * 1{,}000{,}000) / 1{,}000{,}000{,}000) * \text{CGMFC-21: AREA}$$

$$\therefore$$

$$\text{C t per 1 m}^3 = \text{CGMFC-21: AREA} * 0.001 * \text{Bag DT}$$

*C = Carbon, t = metric tons, 1 m is soil depth, CGMFC-21 : AREA is the area of mangrove cover in* $m^2$ *from the CGMFC-21 database, Bag DT is the Jardine and Siikamäki [51] soil carbon measure in mg C per* $mm^2$*. r and P are not possible to construct in the algorithm utilized.*

For our model, we assume a uniform mangrove soil depth of 1 m. This is a conservative estimate of soil carbon stocks because mangrove soil carbon can be found at lower depths [49]. However, our approach is in line with previous global studies[42, 49, 56, 57] and the IPCC[58]. The IPCC's 2013 *Supplement to the 2006 IPCC Guidelines for National Greenhouse Gas Inventories: Wetlands* explicitly recommends that policy makers use a mangrove soil depth of 1 m for calculations of carbon stocks and emissions, under the assumption that in most land use conversions (e.g., to aquaculture) it is the first meter of soil that is most disturbed and most vulnerable to remineralization[56, 58]. Thus, we used this soil depth so that the data provided in this manuscript are as policy-relevant as possible and can be easily incorporated in national emissions reporting mechanisms.

*Calculation of whole-ecosystem carbon stocks and emissions due to deforestation*

We calculated global, national, and regional mangrove carbon stocks by summing all calculated carbon pools: the relevant 123,332,913 individual measures generated from all five equations for mangrove AGB and mangrove BGB, and the pixel-level soil estimates to 1 m depth. Confidence intervals were then calculated at the 95% level for the years 2000 and 2012. To estimate emissions from deforestation, we calculated carbon stocks for 2000 and 2012, based on the global mangrove extent for these years[35]. An inventory change approach was taken to calculate emissions lost between these years, similar to the Tier 1 Approach advocated by the IPCC[58] for emissions calculations.

## Error and Uncertainty

A complete error and uncertainty supplement is provided. Within this summary, we report the input errors reported for all datasets including confidence intervals, significance values, and all other statistical values when reported by the authors for all components of the study. We then account for error in the mangrove area calculation, account for sensitivity in the biomass conversions and soils estimates, and finally conduct a whole system cross-comparison to other available field verified data. We summarize the key findings of the supplement below, but full details can be found in the error and uncertainty supplement.

We assess potential land-cover classification error by re-running a cross-comparison against the only other large -area continuous measure of forest cover available. We find that our mangrove area is within 3.6 % of that contained with the National Land Cover Dataset within the mangrove area of Florida. Additionally, we conducted a literature review across the Americas and compared our mangrove area estimates to other comparable remotely-sensed national mangrove area estimates for similar years. The mangrove area analysis used in this paper and the remotely sensed estimated generated by others, including in-country estimates (Supp. Table 1), are remarkably consistent with an $R^2$ of .97 (n= 29, SE 64441, $P < 0.1$).

For mangrove AGB equations used we generate confidence intervals by processing all data globally for 2000 and 2012. We assess the regional applicability of each equation and produce the best fit model based on all equations utilized. We use equations based not only on latitude but on bioclimatic values and field measures including models known to have differing explanatory potential. All original equation outputs for 2000 and 2012 are freely-distributed in GIS format at full-resolution. The best-fit EQ 5 is distributed for all years between 2000 and 2012 as well as for all countries and all level-one administrative units.

For the conversion between mangrove AGB to mangrove BGB, and between mangrove biomass and carbon, we utilize a mean of the values presented in the academic literature and additionally calculated how sensitive the global estimate is to changes in these values.

For soils, we adjust the soil depth to 2 m and 3 m and demonstrate how this alters the global mangrove carbon forecasts. The 2 m adjustment changes the global mangrove carbon budget from 4.19 Pg to 7.15 Pg while adjusting it to 3m takes it to 10.11 Pg. This emphasizes the importance of soil in the global mangrove carbon budget. Additionally, we present the input soil error measures and discuss in-depth the issues related to estimating soil carbon, particularly accounting for regional variances in the supplemental submission. We make our GIS-data freely available, and it accounts for 1 m, 2 m, 3m soil depths at full-resolution to allow users to adjust for regional soil depth variability based on their region.

Finally, we attempt to capture accumulated error and uncertainty by conducting a cross-comparison of our living biomass data against living mangrove living biomass data for four estuaries in Ecuador for which comparable field-driven estimates of biomass have been generated [43]. We find that mangrove living biomass estimates overlap at the 95% confidence level and have a mean difference of less than 21 t C per ha.

## Open Data

Approximately 1 TB of open raster, vector, and tabular data are posted on the Harvard Dataverse under a CC0 - Public Domain Dedication license that allows full and unrestricted global use of the data generated during this research while giving proper citation to the original author. All error and uncertainty datasets are available in spatial format within the Dataverse accompany this project alongside the mangrove carbon data. The data provided allows for full replication, at the minimum mapping unit, of the results generated during this analysis.

## Methods References

## Supplements

## Error and Uncertainty

The error and uncertainty methodology involves determining the reported sources of input error of each dataset and conducting localized uncertainty analysis based on third-party datasets using other remotely sensed mangrove datasets and field verified data.

The methods developed for this analysis rely on many differing input datasets and models, and each of these datasets and models has their individual level of error and uncertainty. Indeed, the input datasets often rely on other datasets and instruments that have their error. For example, the land cover analysis is derived from remote sensing instruments which have known issues related to atmospheric correction, device error, geolocation error, and numerous other potential opportunities for the introduction of error, uncertainty, and bias [1, 2]. Four of the five carbon equations utilized rely on allometric models, which themselves contain uncertainty and error as well as not explaining the full amount of variability within mangrove AGB. The fifth carbon equation relies on highly generalized and interpolated weather-gauge data that is known to be inconsistently recorded, prone to unit errors (such as the confusion of feet and meters), impacted by highly-localized site conditions (such as mountains, valleys, and urban areas), and the stations themselves are biased in their geographic siting[3].

Not only is the data prone to bias, error, and uncertainty; but the data management process itself likely introduces some error and uncertainty, with the transition from remotely sensed data to higher-level GIS products having the potential to include up to five differing types of error[4]. For example, the above-ground to below-ground biomass conversion only occurs for a central value within a range of possible values that likely differ by species and location. The biomass to carbon conversion again only occurs at a central value within a range that itself is based on data from such laboratory techniques that have their inherent uncertainty and error. To account for the error and uncertainty in the carbon estimates we detail and attempt to quantify error and uncertainty for each step of the process, report all statistical measures of error provided including measures of correlation and statistical uncertainty, and finally conduct a cross-comparison of the data against another published dataset that calculates mangrove carbon using field verified data and continuous mangrove cover.

## Mangrove Area

CGMFC-21[5] is the primary database used for the mangrove area calculation. For the year 2000, the geographic delineation of mangrove is obtained from MFW[6], and the mangrove area within this larger delineation is derived from GFC[7]. For each year after 2000, CGMFC-21 relies on reprocessing of GFC for mangrove forest area calculations. MFW reports a 1one-half pixel error measure but lacks a robust classification error section. GFC reports the accuracy of their tropical classification as 99.7% (n=628, 0.9) [7] and this is the most likely classification within which mangrove will exist.

CGMFC-21 provides a robust error and uncertainty analysis as a supplementary methods submission with an emphasis on omitted sites and reported errors within MFW and GFC. They note that missing sites constitute less than 0.1% of the global mangrove area, and report a cross-comparison assessment conducted within an approximately 1,400 $km^2$ mangrove area within Florida. This cross-comparison uses the National Land Cover Dataset (NLCD)[8, 9]. They find that the 2011 continuous mangrove area reported is within 3.6% of the same area within the NLCD. We reran the same analysis as presented and found an identical result. This is an important finding as no other remotely sensed continuous mangrove forest product appears to exist for a large area to allow for comparison. Finding non-US continuous measures of mangrove canopy cover was not possible as all other data are presence or absence at the pixel level. The full data for this cross-comparison is available in the companion Dataverse.

It is possible to use a rule-of -thumb adjustment to change CGMFC-21 to presence and absence for larger areas such as countries. Although negating one of the major benefits of the paper by generalizing the mangrove area findings, the adjustment does allow for literature comparisons against other nationally remotely sensed estimates of mangrove cover that is all presence or absence based. Within the Americas, we compare adjusted CGMFC-21 mangrove area estimates against regional and in-country remotely sensed estimates. We do this for Brazil, Venezuela, Colombia, Ecuador, Panama, USA, Nicaragua, Honduras, Dominican Republic, Costa Rica, Guatemala, El Salvador, Belize, and Trinidad and Tobago (Supp. Table 1). These countries constitute the majority of the top-50 mangrove holding nations in the Americas[5] with the exceptions of Suriname, Guyana, Mexico, and Cuba for which we could find no suitable comparative data.

Supp. Table 1.

| **Country** | **This Paper (ha) 2000** | **Estimate 1 (ha) 2000** | **Estimate 2 (ha) 2000** | **Estimate 3 (ha) various** | **Estimate 4 (ha) various** |
|---|---|---|---|---|---|
| Brazil | 1,286,886 | 1,063,000[6] | 1,299,947[10] | 1,114,399[11] | |
| Venezuela | 402,640 | 336,000[6] | 356,900[10] | | |
| Colombia | 279,025 | 214,700[6] | 407,926[10] | 407,926[12] | 307,537[13] |
| Panama | 221,310 | 154,304[6] | | 168,677[14] | |
| Ecuador | 156,310 | 136,986[6] | 158,261[10] | 151,920[15] | |
| USA | 267,673 | 236,000[6] | 302,955[10] | 245,257[16] | 240,722[17] |
| Nicaragua | 92,637 | | 67,068[18] | 66,406[19] | |
| Honduras | 88,655 | | | 51,578[20] | |
| Dominican Republic | 16,848 | | | 18,441[21] | |
| Costa Rica | 55,981 | | | 36,153[22] | |
| Guatemala | 44,537 | | | 18,840[23] | |
| El Salvador | 39,434 | | 25,200[18] | 39,160[24] | |
| Belize | 50,779 | | | 74,684[25] | 72,622[26] |
| Trinidad & Tobago | 8,730 | | | 9,369[27] | |

All estimates are remotely sensed, and at least one is an in-country estimate. When dates are not given, they are between 2000 and 2012 and can be found in the source.

The mangrove area analysis used in this paper and the remotely sensed estimated generated by others (Supp. Table 1) are remarkably consistent with an $R^2$ of .97 (n= 29, SE 64441, P < 0.1). Again, although not a validation of the mangrove area dataset utilized, the proximity of these mangrove area values increases confidence in the mangrove estimates on which the carbon estimates are based. Further cross-validation work may be required to see if these findings are duplicated outside of the Americas. The full data for this cross-comparison is available in the companion Dataverse.

## Biomass and Carbon Conversions

Numerous Mangrove AGB to BGB conversion exist and are utilized in the literature [28, 29, 30, 31, 32, 33] and all cluster around the mid-range value of .50 that we used. Conversion factors to convert whole-tree mangrove biomass to mangrove carbon exist in a narrow range of values between 0.45 and 0.50, based on the academic literature [28, 33, 34, 35]. For these calculations, we again selected the mid value of 0.475 to represent the mangrove biomass to mangrove carbon conversion ratio. Although central values were used for both the above-ground to below-ground mangrove biomass conversions and the biomass to carbon conversion we conducted sensitivity analysis at the global scale and adjusted these values to both the minimum reported and the maximum reported values.

## Carbon Equations

Limited error or uncertainty measurements are provided by the authors of EQ 1, the r value is reported in Table 2 in the main paper, and no P value is reported to accompany the r value. Again, limited error measures or uncertainty analysis are provided by the authors of EQ 2, the r value is indicated in Table 2 in the main paper, and no P value is reported to accompany the r value. The authors of EQ 2, additionally report that this model only explains 13.9% of the variance within the data. As with EQ 1 and EQ 2, limited error measures or uncertainty analysis are provided by the authors of EQ 3, both an r value and P value are reported in Table 2 in the main paper. The authors of EQ 4 provide limited error or uncertainty analysis, but they provide an AIC measure of 1.13, an r value reported in Table 2, and no P value to accompany the r value. EQ 4 relies on the bioclimatic variables of Mean Temperature of Warmest Quarter, Mean Temperature of Coldest Quarter, Precipitation of Wettest Quarter, and Precipitation of Driest Quarter in the bioclim database[3]. The error and uncertainty sections of the bioclim database are robust but mostly descriptive in nature. Temperature errors are reported as having mean errors between 0˚ C and 1˚ C but are potentially higher in the Americas[3]. Precipitation cross-comparison is performed, but no single error value is reported aside from noting that the error is greater in regions with higher rainfall and mountainous areas[3]. EQ 5 is a pixel level average off EQ 1 and EQ 2, no independent error assessment is made of this equation beyond noting that both this analysis and a previous analysis[32] have found that the averaging of these latitudinal estimates may provide results more representative of Mangrove AGB than either equation used separately.

## Soil

The choice of a 1 m soil reporting depth is highly subjective although studies have shown this is close to the likely average depth[36], despite adequate inventories not yet existing to establish this figure[28, 36, 37]. This 1 m assumption has become standard in the academic literature[28, 38, 39, 40]. Additionally, we use 1 m to allow for direct comparison of our data with other estimates and to comply with the IPCC finding that it is the top 1 m of the soil is the most vulnerable to remineralization[41, 42]. The 1 m figure is considered a conservative estimate and it is noted that mangroves often exist on carbon rich soils many meters deep, particularly in estuarine systems[28]. Our distributed database is the only product known to us that allows for whole system mangrove carbon to be calculated at 30 m resolution globally using 1 m, 2 m, 3 m, or any combination thereof soil depths. For example, one region can be extracted at 1 m while another can be extracted at 2m, and so on. This variability component matters as regional differences in mangrove soil depth remains an unknown parameter in studies of this nature[28, 36, 37].

Undoubtedly, soil carbon remains the most uncertain of all estimates within the whole system estimates presented in this and other similar paper. Although 900 local measurements feed into the global soil carbon assessment, many important mangrove holding regions such as West Africa appears to be entirely lacking samples, other areas such as South-East India are over-sampled relative to their mangrove holdings, and the samples taken in some regions show a high degree of variability[38]. This variability is important as Figure 1 in the main paper demonstrates that even assuming a 1 m conservative estimate, soils contain approximately 71% of the global whole-system mangrove carbon stocks. This is evidenced by adjusting our 1 m soil estimate to 2 m and assuming no reduction in C at increasing soil depth. This 2 m adjustment changes the global mangrove carbon budget from 4.19 Pg to 7.15 Pg while adjusting it to 3m takes it to 10.11 Pg. The full data for differing soil depths are provided in the companion Dataverse.

The soil data used as input reports ±18.8% as the uncertainty at what appears to be the 95% CI, with an SSE of 0.844 e05 and an average MPE of 32.26 [40]. The machine learning Bag Decision Tree algorithm used in the input soil analysis has not been replicated or applied in any other studies in the academic literature we have observed.

## Cross comparison

Cross comparison is difficult due to the features of this research that make it both unique and novel. For example, few if any continuous mangrove forest cover products exist, where they do exist they have yet to be used to calculate mangrove biomass, living carbon, or whole system mangrove carbon. Where presence or absence mangrove forest cover measures exist, they are often for small study area site estimates and most-often for one year. Finally, what data is distributed for larger area whole-system mangrove carbon is rarely compiled at 30 m or less resolution. For example, the other global estimates are based on climatic data at approximately 1 $km^2$ at the equator[31], gridded into an 81$km^2$ global grid[40], or non-spatial in nature[28]. None of these global mangrove products[28, 31, 40] are not publicly available to other researchers in geospatial format.

A sub-30 m field-research driven estimate of mangrove carbon was produced for 93,452 ha of Ecuador's northern mangrove forests within the provinces of Esmeraldas and Muisné[43]. This analysis was conducted at a sub-grid resolution of 10 m, meaning that although the data is not continuous with 100 measures feeding into every 30 m pixel, they do have nine mangrove measures feeding into each of the 30 m pixels presented ion this paper and these data should be comparable to the findings presented here. No other continuous mangrove product was found globally to allow for cross-comparison that had high enough spatial resolution over a large enough geographic area.

The Ecuadorian researchers estimate mangrove carbon holdings by conducting field research using CIFOR[33] standards for measuring mangrove carbon stocks and applying these data to mangrove stands across four estuaries of northern Ecuador[43]. Importantly, they use a pixel size below 30 m and make their study areas polygons publicly available under a non-commercial license as a supplement to their paper. Their data are for 2011 / 2012 so are also suitable temporally for a cross-comparison. With our findings, The Ecuadorian study only accounts for living carbon, so the soil is not considered in the cross-comparison, but as this comes from a standard database, it should not alter the findings. We use their study areas and then extract out the EQ5-driven mangrove AGB from our database. We then apply both the mangrove AGB to mangrove BGB, and CMB to C, conversions to their study areas but with our data. Our results indicate between 8,275,672 t C and 11,041,416 t C (95% CI) of living mangrove carbon across the entirety of their study areas whereas their field driven method finds between 6,545,157 t C – 8,940,841 t C (95 CI). Supplemental Figure 1 demonstrates the results of the cross-validation.

Supplemental Figure 1

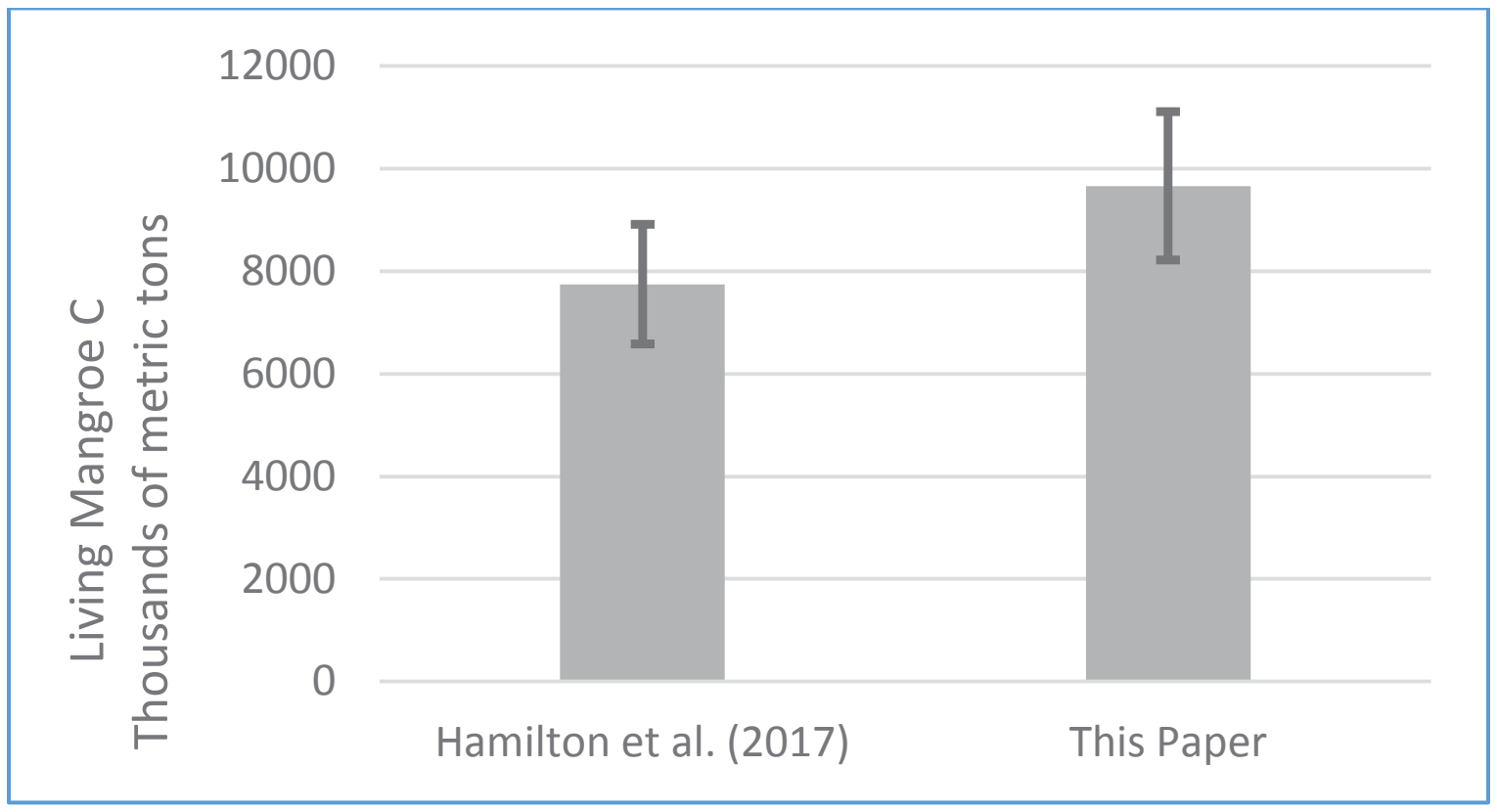

The proximity of this cross-comparison, one based on field collected data at the 10 m level, and the other from our global database increases the confidence of applying our results at the sub-national estuarine level. All cross-comparison data has been made available for verification in out Dataverse account alongside our actual data.

## Error and Uncertainty References

## Supplemental Discussion

The following three reasons likely explain some of the reasons for the lowering of earlier global mangrove carbon estimates. Firstly, Siikamäki, Sanchirico [1] and Jardine and Siikamäki [2] used the MFW database [3] for their estimations. The MFW base map uses a presence or absence approach to mangrove mapping in which each minimum mapping unit has a binary variable depicting mangrove presence or absence, as opposed to 100 continuous measures of canopy cover for each minimum mapping unit used in this analysis. Therefore, the minimum mapping unit used is coded as total mangrove cover when only 75%, 50%, or even 10% of the mapping unit is actually covered by mangrove. This likely causes an over-estimation of mangrove forest cover, especially along the edge of mangrove patches at the site scale, or at the latitudinal limits of mangroves at the regional scales. In both these sites, mangrove boundaries are fuzzy and biomass decreases as mangroves intergrade into other coastal habitats. Hamilton and Casey [4] supplemental appendix S2, provides extensive detail on why such reductions occur and give a mangrove example of why continuous mangrove measures result in approximately a 40% reduction in total area when compared to presence or absence measures. Applying a 40% reduction to the Siikamäki, Sanchirico [1] estimate aligns their global mangrove carbon estimate with the bounds presented in this paper. The continuous cover approach used is this paper is likely an improved estimator of mangrove carbon stocks due to the increase in potential observations at each minimum mapping unit from one to one-hundred. The presence and absence approach may be more suited to analysis were such non-climate related mangrove functions are analyzed such as biodiversity and habitat.

The second reason for differences in the global estimates is likely the coarse 81 $km^2$ grid utilized by Siikamäki, Sanchirico [1], this grid is approximately 90,000 times coarser than the high-resolution grid utilized in this analysis. Aggregating to such large grids may well enforce heterogeneity that may not be present. That is, mangrove canopy cover, soil carbon, bioclimatic variables, and even latitude, may have enough variability within the grid to potentially result in an overestimation of results.

The final reason is that most earlier mangrove global carbon stock estimates integrate the Twilley, Chen [5] latitudinal equation that this research shows may produce global mangrove carbon estimates at the higher end of the latitudinal equation range.

The clustering of the suite of mangrove carbon calculation so tightly around the mean is surprising based on some earlier regional analyses [6, 7]. Such clustering may be partially explained by the models not being truly independent of each other. For example, EQ 1 is actually a subset of EQ 2, and EQ 5 is merely a pixel average of EQ 2 and EQ 3. Even the seemingly independent bioclimatic EQ 4 likely uses some of the same field measures as EQ 1, EQ 2, and EQ 3 during model calibration. Despite this fact, the tight clustering across all five models increases confidence in the estimates provided. No other global analyses use more than one equation so it may just be that the equations tend to cluster around the mean when utilized globally while producing substantially differing regional values. The data produced in this study indicates that this may be the case, particularly when mangroves are in close proximity to the equator.

## Supplemental Discussion References

## Author Contributions

SH developed the equations, designed the study, and carried out the analysis. SH wrote the methods section and created the figures and tables. DF wrote the paper aside from the methods and designed the framework for the paper.

## Competing Financial Interests

Hamilton has no competing financial interests.

Freiss has no competing financial interests.

| Country Name | Mangrove Area (2012) $km^2$ | Mangrove Area Rank | Tonnes of Carbon | | | Percent of Global Total | C Rank | Change |
|---|---|---|---|---|---|---|---|---|
| Indonesia | 23,324.29 | 1 | 1,275,115,175 | ± | 19,597,086 | 30.41 | 1 | 0 |
| Brazil | 7,674.94 | 2 | 389,760,564 | ± | 9,556,539 | 9.30 | 2 | 0 |
| Malaysia | 4,725.84 | 3 | 258,882,085 | ± | 4,002,528 | 6.17 | 3 | 0 |
| Papua New Guinea | 4,172.29 | 4 | 223,096,105 | ± | 3,836,601 | 5.32 | 4 | 0 |
| Australia | 3,316.21 | 5 | 152,539,573 | ± | 2,104,454 | 3.64 | 5 | 0 |
| Mexico | 2,991.83 | 6 | 149,261,592 | ± | 1,203,826 | 3.56 | 6 | 0 |
| Nigeria | 2,653.99 | 7 | 127,914,456 | ± | 2,559,377 | 3.05 | 7 | 0 |
| Myanmar | 2,557.45 | 8 | 118,883,668 | ± | 1,409,261 | 2.84 | 8 | 0 |
| Venezuela | 2,403.83 | 9 | 112,537,865 | ± | 1,851,142 | 2.68 | 9 | 0 |
| Philippines | 2,064.24 | 10 | 104,470,697 | ± | 1,341,367 | 2.49 | 10 | 0 |
| Thailand | 1,886.33 | 11 | 91,793,396 | ± | 1,414,284 | 2.19 | 11 | 0 |
| Colombia | 1,671.86 | 13 | 84,108,157 | ± | 1,831,402 | 2.01 | 12 | 1 |
| Cuba | 1,633.46 | 14 | 81,223,503 | ± | 651,189 | 1.94 | 13 | 1 |
| USA | 1,568.60 | 15 | 75,453,694 | ± | 622,606 | 1.80 | 14 | 1 |
| Bangladesh | 1,772.98 | 12 | 74,049,402 | ± | 653,854 | 1.77 | 15 | -3 |
| Panama | 1,323.94 | 16 | 72,923,978 | ± | 1,222,387 | 1.74 | 16 | 0 |
| Gabon | 1,082.11 | 19 | 58,592,889 | ± | 1,979,216 | 1.40 | 17 | 2 |
| Mozambique | 1,223.67 | 17 | 55,803,315 | ± | 723,403 | 1.33 | 18 | -1 |
| Ecuador | 935.74 | 20 | 55,566,461 | ± | 1,660,042 | 1.33 | 19 | 1 |
| Cameroon | 1,112.76 | 18 | 53,980,215 | ± | 1,138,012 | 1.29 | 20 | -2 |
| Madagascar | | | 40,452,495 | ± | 472,740 | 0.96 | 21 | |
| Guinea | | | 37,970,367 | ± | 521,895 | 0.91 | 22 | |
| India | | | 37,028,410 | ± | 392,240 | 0.88 | 23 | |
| France (Martinique, Guiana, Guadeloupe, Mayotte) | | | 35,950,395 | ± | 591,144 | 0.86 | 24 | |
| Viet Nam | | | 33,825,459 | ± | 630,639 | 0.81 | 25 | |
| Guinea-Bissau | | | 33,517,865 | ± | 532,332 | 0.80 | 26 | |
| Sierra Leone | | | 30,790,720 | ± | 540,394 | 0.73 | 27 | |
| Nicaragua | | | 29,567,060 | ± | 296,163 | 0.71 | 28 | |
| Honduras | | | 28,768,895 | ± | 286,121 | 0.69 | 29 | |
| Suriname | | | 26,827,971 | ± | 442,535 | 0.64 | 30 | |
| Tanzania | | | 23,857,740 | ± | 543,545 | 0.57 | 31 | |
| Solomon Islands | | | 21,522,317 | ± | 492,156 | 0.51 | 32 | |
| Fiji | | | 19,623,559 | ± | 267,631 | 0.47 | 33 | |
| Costa Rica | | | 18,195,395 | ± | 237,135 | 0.43 | 34 | |
| Belize | | | 16,064,627 | ± | 134,401 | 0.38 | 35 | |
| Cambodia | | | 15,540,285 | ± | 195,387 | 0.37 | 36 | |
| Guatemala | | | 13,522,950 | ± | 135,168 | 0.32 | 37 | |
| Kenya | | | 12,601,617 | ± | 370,391 | 0.30 | 38 | |
| El Salvador | | | 11,993,743 | ± | 167,036 | 0.29 | 39 | |
| Guyana | | | 9,905,985 | ± | 132,165 | 0.24 | 40 | |
| Angola | | | 8,814,659 | ± | 302,839 | 0.21 | 41 | |
| Equatorial Guinea | | | 8,217,135 | ± | 241,247 | 0.20 | 42 | |
| Senegal | | | 7,112,574 | ± | 127,883 | 0.17 | 43 | |
| Congo, DR | | | 6,779,725 | ± | 247,222 | 0.16 | 44 | |
| Brunei Darussalam | | | 5,279,849 | ± | 135,099 | 0.13 | 45 | |
| Dominican Republic | | | 5,338,557 | ± | 58,661 | 0.13 | 46 | |
| New Caledonia | | | 4,826,141 | ± | 35,799 | 0.12 | 47 | |
| New Zealand | | | 2,824,781 | ± | 50,234 | 0.07 | 48 | |
| Trinidad and Tobago | | | 2,844,043 | ± | 39,800 | 0.07 | 49 | |
| Palau | | | 2,748,418 | ± | 54,463 | 0.07 | 50 | |

| | | | | | |
|---|---|---|---|---|---|
| Puerto Rico | 2,427,832 | ± | 22,227 | 0.06 | 51 |
| Cayman Islands | 2,327,129 | ± | 19,953 | 0.06 | 52 |
| Puerto Rico | 2,307,087 | ± | 19,871 | 0.06 | 53 |
| Jamaica | 2,291,153 | ± | 17,981 | 0.05 | 54 |
| Haiti | 2,215,778 | ± | 37,642 | 0.05 | 55 |
| Liberia | 2,079,883 | ± | 38,266 | 0.05 | 56 |
| Bahamas | 1,766,687 | ± | 19,335 | 0.04 | 57 |
| Sri Lanka | 1,652,640 | ± | 28,617 | 0.04 | 58 |
| Ghana | 1,103,883 | ± | 31,826 | 0.03 | 59 |
| Côte dIvoire | 958,780 | ± | 25,783 | 0.02 | 60 |
| Peru | 876,842 | ± | 28,186 | 0.02 | 61 |
| Somalia | 612,597 | ± | 14,734 | 0.01 | 62 |
| Vanuatu | 548,263 | ± | 5,159 | 0.01 | 63 |
| Pakistan | 522,082 | ± | 4,229 | 0.01 | 64 |
| Micronesia, FS | 425,290 | ± | 6,148 | 0.01 | 65 |
| Timor-Leste | 404,315 | ± | 9,263 | 0.01 | 66 |
| Benin | 309,695 | ± | 8,409 | 0.01 | 67 |
| China | 289,888 | ± | 15,494 | 0.01 | 68 |
| Seychelles | 287,179 | ± | 6,167 | 0.01 | 69 |
| South Africa | 239,707 | ± | 3,314 | 0.01 | 70 |
| Japan | 73,093 | ± | 8,402 | 0.00 | 71 |
| Singapore | 100,877 | ± | 1,510 | 0.00 | 72 |
| ntigua and Barbuda | 91,190 | ± | 727 | 0.00 | 73 |
| Eritrea | 83,147 | ± | 797 | 0.00 | 74 |
| /irgin Islands, U.S. | 65,407 | ± | 557 | 0.00 | 75 |
| Grenada | 64,405 | ± | 626 | 0.00 | 76 |
| Saint Lucia | 61,540 | ± | 579 | 0.00 | 77 |
| etherlands - Bonaire | 60,765 | ± | 790 | 0.00 | 78 |
| Comoros | 37,553 | ± | 463 | 0.00 | 79 |
| g Kong (SAR, China) | 27,537 | ± | 1,380 | 0.00 | 80 |
| Togo | 26,905 | ± | 772 | 0.00 | 81 |
| Curaçao | 18,454 | ± | 262 | 0.00 | 82 |
| Maldives | 16,751 | ± | 295 | 0.00 | 83 |
| Saudi Arabia | 14,824 | ± | 125 | 0.00 | 84 |
| rgin Islands, British | 13,658 | ± | 120 | 0.00 | 85 |
| nited Arab Emirates | 11,461 | ± | 94 | 0.00 | 86 |
| Vincent and Grenadines | 10,907 | ± | 106 | 0.00 | 87 |
| Turks and Caicos | 8,355 | ± | 60 | 0.00 | 88 |
| Djibouti | 8,084 | ± | 128 | 0.00 | 89 |
| Taiwan | 6,983 | ± | 392 | 0.00 | 90 |
| Aruba | 6,817 | ± | 94 | 0.00 | 91 |
| aint Kitts and Nevis | 6,725 | ± | 59 | 0.00 | 92 |
| Yemen | 6,685 | ± | 87 | 0.00 | 93 |
| Somaliland | 4,649 | ± | 58 | 0.00 | 94 |
| Barbados | 3,217 | ± | 39 | 0.00 | 95 |
| Morocco | 1,864 | ± | 189 | 0.00 | 96 |
| Anguilla | 1,935 | ± | 17 | 0.00 | 97 |
| Saint Martin | 1,807 | ± | 17 | 0.00 | 98 |
| Iran | 1,524 | ± | 13 | 0.00 | 99 |
| Oman | 892 | ± | 7 | 0.00 | 100 |
| Sudan | 626 | ± | 6 | 0.00 | 101 |
| Bermuda | 146 | ± | 27 | 0.00 | 102 |
| Egypt | 240 | ± | 3 | 0.00 | 103 |
| Qatar | 46 | ± | 1 | 0.00 | 104 |
| Saint Barthélemy | 44 | ± | 1 | 0.00 | 105 |
| acao (SAR, China) | 15 | ± | 1 | 0.00 | 106 |
| Mauritania | 1 | ± | 0 | 0.00 | 107 |
| **TOTAL** | **4,192,756,365** | | | **100** | |

| Country | Level One Name | C t 2012 | % of Global Total | Running % of Global Total |
|---|---|---|---|---|
| Indonesia | Papua | 328,816,690 | 7.77% | 7.77% |
| Indonesia | Irian Jaya Barat | 237,459,220 | 5.61% | 13.38% |
| Brazil | Maranhão | 155,013,142 | 3.66% | 17.04% |
| Malaysia | Sabah | 137,359,199 | 3.25% | 20.29% |
| Papua New Guinea | Gulf | 122,124,709 | 2.89% | 23.17% |
| Brazil | Pará | 106,739,631 | 2.52% | 25.70% |
| Indonesia | Kalimantan Timur | 95,815,540 | 2.26% | 27.96% |
| Indonesia | Maluku | 92,862,422 | 2.19% | 30.16% |
| Myanmar | Tanintharyi | 87,519,738 | 2.07% | 32.22% |
| Indonesia | Sumatera Selatan | 81,738,975 | 1.93% | 34.15% |
| Australia | Queensland | 80,763,628 | 1.91% | 36.06% |
| Brazil | Amapá | 76,341,229 | 1.80% | 37.87% |
| United States | Florida | 74,257,077 | 1.75% | 39.62% |
| Indonesia | Riau | 74,093,223 | 1.75% | 41.37% |
| Malaysia | Sarawak | 73,613,945 | 1.74% | 43.11% |
| Bangladesh | Khulna | 66,020,389 | 1.56% | 44.67% |
| Indonesia | Kalimantan Utara | 63,409,405 | 1.50% | 46.17% |
| Indonesia | Kalimantan Barat | 63,251,409 | 1.49% | 47.66% |
| Mexico | Campeche | 55,642,333 | 1.31% | 48.98% |
| Australia | Northern Territory | 53,906,863 | 1.27% | 50.25% |
| Nigeria | Bayelsa | 52,005,529 | 1.23% | 51.48% |
| Venezuela | Delta Amacuro | 49,813,122 | 1.18% | 52.66% |
| Gabon | Estuaire | 44,296,888 | 1.05% | 53.71% |
| Nigeria | Delta | 40,787,194 | 0.96% | 54.67% |
| Ecuador | Guayas | 40,080,774 | 0.95% | 55.62% |
| Colombia | Nariño | 37,417,294 | 0.88% | 56.50% |
| Cameroon | Sud-Ouest | 32,592,081 | 0.77% | 57.27% |
| French Guiana | Cayenne | 32,093,971 | 0.76% | 58.03% |
| Indonesia | Bangka-Belitung | 31,696,460 | 0.75% | 58.78% |
| Venezuela | Monagas | 30,261,030 | 0.72% | 59.49% |
| Indonesia | Sulawesi Tenggara | 28,740,685 | 0.68% | 60.17% |
| Indonesia | Kalimantan Selatan | 27,612,225 | 0.65% | 60.82% |
| Papua New Guinea | Central | 27,442,513 | 0.65% | 61.47% |
| Nigeria | Rivers | 25,801,327 | 0.61% | 62.08% |
| India | Andaman and Nicobar | 25,733,933 | 0.61% | 62.69% |
| Papua New Guinea | Western | 25,531,358 | 0.60% | 63.29% |
| Indonesia | Kepulauan Riau | 25,320,951 | 0.60% | 63.89% |
| Venezuela | Sucre | 24,282,755 | 0.57% | 64.47% |
| Philippines | Palawan | 23,946,664 | 0.57% | 65.03% |
| Sierra Leone | Southern | 23,864,629 | 0.56% | 65.60% |
| Mozambique | Zambezia | 22,767,536 | 0.54% | 66.13% |
| Myanmar | Rakhine | 22,680,634 | 0.54% | 66.67% |
| Cameroon | Littoral | 22,031,951 | 0.52% | 67.19% |
| Madagascar | Mahajanga | 21,729,623 | 0.51% | 67.70% |
| Guinea | Boké | 21,701,887 | 0.51% | 68.22% |
| Indonesia | Maluku Utara | 21,465,426 | 0.51% | 68.72% |
| Brazil | Bahia | 21,426,224 | 0.51% | 69.23% |
| Indonesia | Kalimantan Tengah | 20,631,042 | 0.49% | 69.72% |
| Cuba | Pinar del Río | 20,313,938 | 0.48% | 70.20% |
| Panama | Darién | 19,874,894 | 0.47% | 70.67% |
| Thailand | Phangnga | 19,542,821 | 0.46% | 71.13% |
| Mexico | Chiapas | 19,105,754 | 0.45% | 71.58% |
| Indonesia | Sumatera Utara | 18,482,110 | 0.44% | 72.02% |
| Nicaragua | Atlántico Norte | 17,920,229 | 0.42% | 72.44% |
| Mexico | Tabasco | 17,586,906 | 0.42% | 72.86% |
| Malaysia | Perak | 17,040,251 | 0.40% | 73.26% |
| Gabon | Ogooué-Maritime | 16,074,956 | 0.38% | 73.64% |
| Thailand | Trang | 15,480,756 | 0.37% | 74.00% |
| Panama | Chiriquí | 15,362,599 | 0.36% | 74.37% |
| Indonesia | Sulawesi Tengah | 14,991,304 | 0.35% | 74.72% |
| Honduras | Gracias a Dios | 14,959,498 | 0.35% | 75.07% |
| Mexico | Veracruz | 14,732,205 | 0.35% | 75.42% |
| Panama | Veraguas | 14,548,415 | 0.34% | 75.77% |
| Costa Rica | Puntarenas | 14,058,607 | 0.33% | 76.10% |
| Guinea | Kindia | 13,886,987 | 0.33% | 76.43% |
| Papua New Guinea | Milne Bay | 13,798,902 | 0.33% | 76.75% |
| Mozambique | Sofala | 13,698,559 | 0.32% | 77.08% |
| Australia | Western Australia | 13,466,920 | 0.32% | 77.39% |
| Thailand | Satun | 13,249,340 | 0.31% | 77.71% |
| Vietnam | Cà Mau | 13,150,216 | 0.31% | 78.02% |
| Thailand | Krabi | 13,013,402 | 0.31% | 78.33% |
| Vietnam | Hồ Chí Minh city | 12,428,441 | 0.29% | 78.62% |
| Mexico | Yucatán | 12,397,672 | 0.29% | 78.91% |
| Cambodia | Kaôh Kong | 12,369,829 | 0.29% | 79.20% |
| Tanzania | Pwani | 11,864,706 | 0.28% | 79.49% |
| Cuba | Camagüey | 11,528,138 | 0.27% | 79.76% |
| Madagascar | Antsiranana | 11,292,823 | 0.27% | 80.02% |
| Panama | Panamá | 11,176,862 | 0.26% | 80.29% |
| Malaysia | Johor | 10,966,028 | 0.26% | 80.55% |
| Colombia | Valle del Cauca | 10,934,216 | 0.26% | 80.81% |
| Cuba | Matanzas | 10,751,333 | 0.25% | 81.06% |
| Belize | Belize | 10,451,637 | 0.25% | 81.31% |
| India | West Bengal | 10,218,281 | 0.24% | 81.55% |
| Guinea-Bissau | Cacheu | 9,963,981 | 0.24% | 81.78% |
| Ecuador | Esmeraldas | 9,800,671 | 0.23% | 82.02% |
| Cuba | Villa Clara | 9,713,293 | 0.23% | 82.25% |
| Suriname | Saramacca | 9,429,266 | 0.22% | 82.47% |
| Philippines | Sulu | 9,315,415 | 0.22% | 82.69% |
| Guinea-Bissau | Tombali | 9,189,709 | 0.22% | 82.91% |
| Colombia | Cauca | 9,186,397 | 0.22% | 83.12% |
| Equatorial Guinea | Litoral | 8,895,305 | 0.21% | 83.33% |
| Indonesia | Jawa Timur | 8,813,712 | 0.21% | 83.54% |
| Suriname | Commewijne | 8,686,157 | 0.21% | 83.75% |
| Colombia | Chocó | 8,683,638 | 0.21% | 83.95% |
| Thailand | Ranong | 8,651,682 | 0.20% | 84.16% |
| Malaysia | Selangor | 8,643,641 | 0.20% | 84.36% |
| Nicaragua | Chinandega | 8,630,945 | 0.20% | 84.56% |
| Mozambique | Nampula | 8,439,722 | 0.20% | 84.76% |
| Indonesia | Aceh | 8,396,909 | 0.20% | 84.96% |
| Papua New Guinea | Oro | 8,333,632 | 0.20% | 85.16% |
| Indonesia | Sumatera Barat | 8,101,389 | 0.19% | 85.35% |
| Kenya | Lamu | 8,012,445 | 0.19% | 85.54% |
| Myanmar | Ayeyarwady | 7,977,126 | 0.19% | 85.73% |
| Solomon Islands | Isabel | 7,970,478 | 0.19% | 85.92% |
| Papua New Guinea | New Ireland | 7,858,352 | 0.19% | 86.10% |
| Brazil | Paraná | 7,748,948 | 0.18% | 86.28% |
| Madagascar | Toliary | 7,729,876 | 0.18% | 86.47% |
| Guinea-Bissau | Bolama | 7,491,925 | 0.18% | 86.64% |
| Mozambique | Cabo Delgado | 7,345,972 | 0.17% | 86.82% |
| Indonesia | Sulawesi Selatan | 7,322,074 | 0.17% | 86.99% |
| Bangladesh | Barisal | 7,288,992 | 0.17% | 87.16% |
| Venezuela | Zulia | 7,117,199 | 0.17% | 87.33% |
| Democratic Republic of the Congo | Bas-Congo | 6,970,424 | 0.16% | 87.50% |
| Papua New Guinea | West New Britain | 6,965,453 | 0.16% | 87.66% |
| Angola | Zaire | 6,903,435 | 0.16% | 87.82% |
| Fiji | Northern | 6,860,220 | 0.16% | 87.99% |
| Guyana | Barima-Waini | 6,832,064 | 0.16% | 88.15% |
| Mexico | Quintana Roo | 6,760,541 | 0.16% | 88.31% |
| Mexico | Nayarit | 6,757,261 | 0.16% | 88.47% |
| El Salvador | Usulután | 6,553,008 | 0.15% | 88.62% |
| Fiji | Central | 6,405,877 | 0.15% | 88.77% |
| Sierra Leone | Northern | 6,371,898 | 0.15% | 88.92% |
| Guatemala | Izabal | 6,152,659 | 0.15% | 89.07% |
| Ecuador | El Oro | 6,018,578 | 0.14% | 89.21% |
| Indonesia | Nusa Tenggara Timur | 5,973,290 | 0.14% | 89.35% |
| Nigeria | Cross River | 5,877,854 | 0.14% | 89.49% |
| Mexico | Oaxaca | 5,779,275 | 0.14% | 89.63% |
| Brazil | São Paulo | 5,776,719 | 0.14% | 89.76% |
| Honduras | Valle | 5,774,227 | 0.14% | 89.90% |
| Cuba | Isla de la Juventud | 5,743,340 | 0.14% | 90.04% |
| Indonesia | Sulawesi Utara | 5,559,599 | 0.13% | 90.17% |
| Philippines | Quezon | 5,442,173 | 0.13% | 90.30% |
| Philippines | Zamboanga Sibugay | 5,421,163 | 0.13% | 90.42% |
| Fiji | Western | 5,374,385 | 0.13% | 90.55% |
| Colombia | Magdalena | 5,350,128 | 0.13% | 90.68% |
| Philippines | Tawi-Tawi | 5,305,059 | 0.13% | 90.80% |
| Thailand | Nakhon Si Thammarat | 5,260,701 | 0.12% | 90.93% |
| Philippines | Surigao del Norte | 5,119,534 | 0.12% | 91.05% |
| Solomon Islands | Malaita | 4,948,449 | 0.12% | 91.17% |
| Cuba | Ciego de Ávila | 4,869,136 | 0.12% | 91.28% |
| Tanzania | Lindi | 4,664,974 | 0.11% | 91.39% |
| Indonesia | Gorontalo | 4,616,936 | 0.11% | 91.50% |
| Philippines | Samar | 4,524,947 | 0.11% | 91.61% |
| Colombia | Atlántico | 4,381,456 | 0.10% | 91.71% |
| Brazil | Pernambuco | 4,371,835 | 0.10% | 91.81% |
| Costa Rica | Guanacaste | 4,277,045 | 0.10% | 91.91% |
| Mexico | Sinaloa | 4,255,999 | 0.10% | 92.02% |
| Brazil | Sergipe | 4,246,664 | 0.10% | 92.12% |
| Brunei | Temburong | 4,203,510 | 0.10% | 92.21% |
| Senegal | Ziguinchor | 4,172,162 | 0.10% | 92.31% |
| Thailand | Trat | 4,119,379 | 0.10% | 92.41% |
| Philippines | Bohol | 4,075,394 | 0.10% | 92.51% |
| Guinea-Bissau | Quinara | 3,956,319 | 0.09% | 92.60% |
| Philippines | Zamboanga del Sur | 3,767,438 | 0.09% | 92.69% |
| Cuba | Holguín | 3,723,294 | 0.09% | 92.78% |
| Brazil | Paraíba | 3,692,279 | 0.09% | 92.86% |
| Honduras | Choluteca | 3,662,251 | 0.09% | 92.95% |
| Belize | Stann Creek | 3,595,778 | 0.08% | 93.04% |
| Suriname | Nickerie | 3,532,543 | 0.08% | 93.12% |
| Philippines | Basilan | 3,531,357 | 0.08% | 93.20% |
| Solomon Islands | Western | 3,473,928 | 0.08% | 93.29% |
| Malaysia | Kedah | 3,389,446 | 0.08% | 93.37% |
| Cuba | Granma | 3,388,008 | 0.08% | 93.45% |
| Nigeria | Akwa Ibom | 3,279,299 | 0.08% | 93.52% |
| Suriname | Marowijne | 3,230,282 | 0.08% | 93.60% |
| Thailand | Chanthaburi | 3,150,574 | 0.07% | 93.67% |
| Australia | South Australia | 3,115,757 | 0.07% | 93.75% |
| Indonesia | Bali | 2,999,106 | 0.07% | 93.82% |
| Senegal | Fatick | 2,992,007 | 0.07% | 93.89% |
| Mozambique | Inhambane | 2,976,393 | 0.07% | 93.96% |
| Guatemala | Retalhuleu | 2,926,377 | 0.07% | 94.03% |
| Cuba | Sancti Spíritus | 2,878,575 | 0.07% | 94.10% |
| Philippines | Eastern Samar | 2,849,089 | 0.07% | 94.16% |
| Panama | Los Santos | 2,801,518 | 0.07% | 94.23% |
| Kenya | Kwale | 2,771,824 | 0.07% | 94.30% |
| Cambodia | Krong Preah Sihanouk | 2,763,946 | 0.07% | 94.36% |
| Brazil | Rio Grande do Norte | 2,748,228 | 0.06% | 94.43% |
| Honduras | Colón | 2,703,499 | 0.06% | 94.49% |
| Papua New Guinea | Manus | 2,692,486 | 0.06% | 94.55% |
| Brazil | Rio de Janeiro | 2,688,651 | 0.06% | 94.62% |
| Indonesia | Jawa Tengah | 2,687,136 | 0.06% | 94.68% |
| Brazil | Santa Catarina | 2,656,196 | 0.06% | 94.74% |
| Tanzania | Tanga | 2,643,653 | 0.06% | 94.81% |
| Colombia | Córdoba | 2,625,266 | 0.06% | 94.87% |
| New Caledonia | Nord | 2,621,576 | 0.06% | 94.93% |
| Cuba | Mayabeque | 2,611,233 | 0.06% | 94.99% |
| French Guiana | Saint-Laurent-du-Maroni | 2,586,793 | 0.06% | 95.05% |
| Cuba | Las Tunas | 2,579,472 | 0.06% | 95.11% |
| Panama | Panamá Oeste | 2,546,782 | 0.06% | 95.17% |
| Philippines | Leyte | 2,465,688 | 0.06% | 95.23% |
| El Salvador | La Paz | 2,443,591 | 0.06% | 95.29% |
| Philippines | Surigao del Sur | 2,377,267 | 0.06% | 95.35% |
| Guinea-Bissau | Biombo | 2,364,493 | 0.06% | 95.40% |
| Brazil | Espírito Santo | 2,356,509 | 0.06% | 95.46% |
| Panama | Coclé | 2,341,571 | 0.06% | 95.51% |
| Colombia | Sucre | 2,332,139 | 0.06% | 95.57% |
| Papua New Guinea | Bougainville | 2,281,134 | 0.05% | 95.62% |
| Malaysia | Trengganu | 2,269,027 | 0.05% | 95.67% |
| New Caledonia | Sud | 2,193,741 | 0.05% | 95.73% |
| Papua New Guinea | East Sepik | 2,189,734 | 0.05% | 95.78% |
| Suriname | Coronie | 2,122,439 | 0.05% | 95.83% |
| Indonesia | Lampung | 2,110,038 | 0.05% | 95.88% |
| Tanzania | Mtwara | 2,104,328 | 0.05% | 95.93% |
| Brazil | Ceará | 2,089,179 | 0.05% | 95.98% |
| Panama | Kuna Yala | 2,074,496 | 0.05% | 96.03% |
| Solomon Islands | Choiseul | 2,030,600 | 0.05% | 96.07% |
| Thailand | Chumphon | 2,029,203 | 0.05% | 96.12% |
| Guinea | Conakry | 1,972,675 | 0.05% | 96.17% |
| Vietnam | Đồng Nai | 1,949,470 | 0.05% | 96.22% |
| Philippines | Northern Samar | 1,914,439 | 0.05% | 96.26% |
| Nicaragua | León | 1,891,938 | 0.04% | 96.31% |
| Philippines | Camarines Sur | 1,853,691 | 0.04% | 96.35% |
| Australia | New South Wales | 1,834,601 | 0.04% | 96.39% |
| Philippines | Cagayan | 1,823,720 | 0.04% | 96.44% |
| Mexico | Guerrero | 1,809,358 | 0.04% | 96.48% |
| El Salvador | La Unión | 1,808,581 | 0.04% | 96.52% |
| Philippines | Masbate | 1,755,343 | 0.04% | 96.56% |
| Thailand | Surat Thani | 1,742,448 | 0.04% | 96.60% |
| Guatemala | Escuintla | 1,735,318 | 0.04% | 96.64% |
| Panama | Bocas del Toro | 1,708,546 | 0.04% | 96.68% |
| Cayman Islands | North Side | 1,670,083 | 0.04% | 96.72% |
| Thailand | Pattani | 1,654,779 | 0.04% | 96.76% |
| Indonesia | Sulawesi Barat | 1,646,720 | 0.04% | 96.80% |
| Trinidad and Tobago | San Juan-Laventille | 1,573,911 | 0.04% | 96.84% |
| Guyana | Pomeroon-Supenaam | 1,558,426 | 0.04% | 96.88% |
| Kenya | Kilifi | 1,526,514 | 0.04% | 96.91% |
| New Zealand | Northland | 1,524,177 | 0.04% | 96.95% |
| Colombia | Bolívar | 1,514,495 | 0.04% | 96.98% |
| Angola | Luanda | 1,513,371 | 0.04% | 97.02% |
| Papua New Guinea | Morobe | 1,497,285 | 0.04% | 97.06% |
| Mexico | Tamaulipas | 1,472,681 | 0.03% | 97.09% |
| Philippines | Sorsogon | 1,460,287 | 0.03% | 97.12% |
| Guatemala | Santa Rosa | 1,442,690 | 0.03% | 97.16% |
| Venezuela | Miranda | 1,435,390 | 0.03% | 97.19% |
| Vietnam | Kiên Giang | 1,434,347 | 0.03% | 97.23% |
| Philippines | Negros Occidental | 1,419,012 | 0.03% | 97.26% |
| Guinea-Bissau | Oio | 1,405,175 | 0.03% | 97.29% |
| Malaysia | Pahang | 1,395,374 | 0.03% | 97.33% |
| Indonesia | Nusa Tenggara Barat | 1,335,080 | 0.03% | 97.36% |
| Cuba | La Habana | 1,332,890 | 0.03% | 97.39% |
| Tanzania | Pemba North | 1,300,657 | 0.03% | 97.42% |
| Indonesia | Jambi | 1,282,491 | 0.03% | 97.45% |
| Brazil | Alagoas | 1,262,805 | 0.03% | 97.48% |
| Dominican Republic | Samaná | 1,245,326 | 0.03% | 97.51% |
| Peru | Tumbes | 1,209,617 | 0.03% | 97.54% |
| Vietnam | Bà Rịa - Vũng Tàu | 1,169,973 | 0.03% | 97.57% |
| Dominican Republic | Monte Cristi | 1,167,135 | 0.03% | 97.59% |
| Philippines | Marinduque | 1,148,448 | 0.03% | 97.62% |
| Philippines | Cebu | 1,129,291 | 0.03% | 97.65% |
| Belize | Corozal | 1,117,917 | 0.03% | 97.67% |
| Malaysia | Kelantan | 1,104,460 | 0.03% | 97.70% |
| Solomon Islands | Temotu | 1,098,074 | 0.03% | 97.73% |
| Brazil | Piauí | 1,087,727 | 0.03% | 97.75% |
| Philippines | Camarines Norte | 1,078,458 | 0.03% | 97.78% |
| Philippines | Oriental Mindoro | 1,073,198 | 0.03% | 97.80% |
| Mexico | Baja California Sur | 1,062,925 | 0.03% | 97.83% |
| Sierra Leone | Western | 1,054,579 | 0.02% | 97.85% |
| Papua New Guinea | East New Britain | 1,041,131 | 0.02% | 97.88% |
| Panama | Colón | 1,022,765 | 0.02% | 97.90% |
| Nigeria | Lagos | 1,004,290 | 0.02% | 97.92% |
| Nicaragua | Atlántico Sur | 1,002,515 | 0.02% | 97.95% |
| Solomon Islands | Central | 999,428 | 0.02% | 97.97% |
| Vietnam | Quảng Ninh | 992,663 | 0.02% | 98.00% |
| Philippines | Misamis Occidental | 965,426 | 0.02% | 98.02% |
| Indonesia | Banten | 949,003 | 0.02% | 98.04% |
| Philippines | Davao Oriental | 938,998 | 0.02% | 98.06% |
| Gambia | North Bank | 935,846 | 0.02% | 98.09% |
| Belize | Toledo | 935,644 | 0.02% | 98.11% |
| Brunei | Brunei and Muara | 921,760 | 0.02% | 98.13% |
| Nigeria | Edo | 918,492 | 0.02% | 98.15% |
| Honduras | Atlántida | 907,041 | 0.02% | 98.17% |
| Colombia | Antioquia | 905,746 | 0.02% | 98.19% |
| Thailand | Samut Songkhram | 901,654 | 0.02% | 98.21% |
| Mozambique | Maputo | 893,499 | 0.02% | 98.24% |
| Mexico | Jalisco | 833,994 | 0.02% | 98.26% |
| Thailand | Phuket | 831,644 | 0.02% | 98.28% |
| Venezuela | Falcón | 823,235 | 0.02% | 98.29% |
| Tanzania | Pemba South | 796,996 | 0.02% | 98.31% |
| Fiji | Eastern | 790,893 | 0.02% | 98.33% |
| Gambia | Western | 790,305 | 0.02% | 98.35% |
| Thailand | Samut Sakhon | 772,974 | 0.02% | 98.37% |
| Philippines | Lanao del Norte | 751,682 | 0.02% | 98.39% |
| Ecuador | Manabi | 743,534 | 0.02% | 98.40% |
| New Zealand | Auckland | 741,840 | 0.02% | 98.42% |
| Guadeloupe | Pointe-à-Pitre | 739,474 | 0.02% | 98.44% |
| El Salvador | Ahuachapán | 732,035 | 0.02% | 98.46% |
| Philippines | Maguindanao | 724,241 | 0.02% | 98.47% |
| Mexico | Colima | 714,252 | 0.02% | 98.49% |
| Malaysia | Pulau Pinang | 704,998 | 0.02% | 98.51% |
| Philippines | Negros Oriental | 702,293 | 0.02% | 98.52% |
| Ecuador | Galápagos | 699,050 | 0.02% | 98.54% |
| Vietnam | Sóc Trăng | 698,485 | 0.02% | 98.56% |
| Philippines | Zamboanga del Norte | 694,408 | 0.02% | 98.57% |
| Vietnam | Bến Tre | 689,406 | 0.02% | 98.59% |
| Guyana | East Berbice-Corentyne | 687,665 | 0.02% | 98.61% |
| Panama | Herrera | 682,939 | 0.02% | 98.62% |
| Liberia | GrandBassa | 678,436 | 0.02% | 98.64% |
| Cuba | Cienfuegos | 674,266 | 0.02% | 98.65% |
| Nigeria | Ondo | 667,915 | 0.02% | 98.67% |
| Philippines | Dinagat Islands | 667,176 | 0.02% | 98.69% |
| Papua New Guinea | Madang | 656,737 | 0.02% | 98.70% |
| Malaysia | Negeri Sembilan | 654,282 | 0.02% | 98.72% |
| Puerto Rico | Loíza | 652,415 | 0.02% | 98.73% |
| Ghana | Volta | 650,515 | 0.02% | 98.75% |
| India | Andhra Pradesh | 642,333 | 0.02% | 98.76% |
| Somalia | Jubbada Hoose | 626,478 | 0.01% | 98.78% |
| Haiti | L'Artibonite | 605,662 | 0.01% | 98.79% |
| Philippines | Capiz | 581,744 | 0.01% | 98.81% |
| Philippines | Catanduanes | 575,381 | 0.01% | 98.82% |
| Philippines | Occidental Mindoro | 573,907 | 0.01% | 98.83% |
| Tanzania | Zanzibar South and Central | 568,817 | 0.01% | 98.85% |
| Haiti | Nord | 562,811 | 0.01% | 98.86% |
| Angola | Bengo | 548,527 | 0.01% | 98.87% |
| Indonesia | Jawa Barat | 547,726 | 0.01% | 98.89% |
| Côte d'Ivoire | Bas-Sassandra | 540,317 | 0.01% | 98.90% |
| Thailand | Phetchaburi | 536,867 | 0.01% | 98.91% |
| Thailand | Songkhla | 534,515 | 0.01% | 98.92% |
| Dominican Republic | La Altagracia | 524,063 | 0.01% | 98.94% |
| Pakistan | Sind | 520,852 | 0.01% | 98.95% |
| Trinidad and Tobago | Siparia | 518,662 | 0.01% | 98.96% |
| Myanmar | Mon | 517,825 | 0.01% | 98.97% |
| Palau | Airai | 513,759 | 0.01% | 98.98% |
| Guadeloupe | Basse-Terre | 509,820 | 0.01% | 99.00% |
| Honduras | Islas de la Bahía | 508,585 | 0.01% | 99.01% |
| Thailand | Rayong | 481,978 | 0.01% | 99.02% |
| Vanuatu | Malampa | 477,615 | 0.01% | 99.03% |
| Cuba | Guantánamo | 473,042 | 0.01% | 99.04% |
| Dominican Republic | Pedernales | 470,593 | 0.01% | 99.05% |
| Dominican Republic | Hato Mayor | 470,077 | 0.01% | 99.06% |
| Guyana | Mahaica-Berbice | 467,874 | 0.01% | 99.08% |
| Kenya | Mombasa | 457,694 | 0.01% | 99.09% |
| Solomon Islands | Guadalcanal | 456,223 | 0.01% | 99.10% |
| Philippines | Sultan Kudarat | 454,869 | 0.01% | 99.11% |
| Jamaica | Saint Catherine | 453,843 | 0.01% | 99.12% |
| Guatemala | Jutiapa | 445,985 | 0.01% | 99.13% |
| Dominican Republic | El Seybo | 434,630 | 0.01% | 99.14% |
| Philippines | Albay | 431,218 | 0.01% | 99.15% |
| Côte d'Ivoire | Comoé | 425,396 | 0.01% | 99.16% |
| Micronesia | Yap | 423,031 | 0.01% | 99.17% |
| Australia | Victoria | 417,012 | 0.01% | 99.18% |
| New Zealand | Waikato | 416,264 | 0.01% | 99.19% |
| Gabon | Nyanga | 397,426 | 0.01% | 99.20% |
| Cameroon | Sud | 396,647 | 0.01% | 99.21% |
| Jamaica | Westmoreland | 396,354 | 0.01% | 99.22% |
| Guatemala | Suchitepéquez | 392,933 | 0.01% | 99.23% |
| Jamaica | Clarendon | 392,502 | 0.01% | 99.24% |
| Jamaica | Saint Thomas | 389,175 | 0.01% | 99.25% |
| El Salvador | San Vicente | 386,415 | 0.01% | 99.25% |
| Philippines | Aklan | 385,264 | 0.01% | 99.26% |
| Cayman Islands | East End | 383,968 | 0.01% | 99.27% |
| Guatemala | San Marcos | 379,602 | 0.01% | 99.28% |
| Suriname | Paramaribo | 370,995 | 0.01% | 99.29% |
| Tanzania | Dar es Salaam | 370,994 | 0.01% | 99.30% |
| Sri Lanka | Puttalam | 353,817 | 0.01% | 99.31% |
| Philippines | Iloilo | 349,946 | 0.01% | 99.32% |
| Dominican Republic | Puerto Plata | 341,509 | 0.01% | 99.32% |
| Cambodia | Kâmpôt | 341,346 | 0.01% | 99.33% |
| Haiti | Nord-Est | 336,883 | 0.01% | 99.34% |
| Philippines | Antique | 328,893 | 0.01% | 99.35% |
| Gambia | Lower River | 318,918 | 0.01% | 99.36% |
| Liberia | Grand Cape Mount | 316,525 | 0.01% | 99.36% |
| Liberia | River Cess | 316,247 | 0.01% | 99.37% |
| Bangladesh | Chittagong | 315,781 | 0.01% | 99.38% |
| Dominican Republic | Espaillat | 311,985 | 0.01% | 99.39% |
| India | Maharashtra | 310,113 | 0.01% | 99.39% |
| Palau | Peleliu | 307,391 | 0.01% | 99.40% |
| Martinique | Le Marin | 306,612 | 0.01% | 99.41% |
| Haiti | Nippes | 306,456 | 0.01% | 99.41% |
| Vietnam | Trà Vinh | 306,445 | 0.01% | 99.42% |
| Malaysia | Melaka | 306,321 | 0.01% | 99.43% |
| Vietnam | Quảng Bình | 300,208 | 0.01% | 99.44% |
| Seychelles | Outer Islands | 293,659 | 0.01% | 99.44% |
| Thailand | Samut Prakan | 290,323 | 0.01% | 99.45% |
| Palau | Ngeremlengui | 288,168 | 0.01% | 99.46% |
| Palau | Ngaraard | 284,056 | 0.01% | 99.46% |
| El Salvador | Sonsonate | 282,426 | 0.01% | 99.47% |
| Vietnam | Nam Định | 278,084 | 0.01% | 99.48% |
| Thailand | Chachoengsao | 272,536 | 0.01% | 99.48% |
| Philippines | Romblon | 271,473 | 0.01% | 99.49% |
| Angola | Cuanza Sul | 266,696 | 0.01% | 99.50% |
| Philippines | Southern Leyte | 263,340 | 0.01% | 99.50% |
| Palau | Aimeliik | 260,437 | 0.01% | 99.51% |
| Solomon Islands | Makira Ulawa | 258,487 | 0.01% | 99.51% |
| Sri Lanka | Batticaloa | 255,692 | 0.01% | 99.52% |
| Vietnam | Thái Bình | 254,224 | 0.01% | 99.53% |
| Jamaica | Saint Elizabeth | 250,789 | 0.01% | 99.53% |
| Philippines | Lanao del Sur | 246,324 | 0.01% | 99.54% |
| Philippines | Zambales | 245,893 | 0.01% | 99.54% |
| Kenya | Tana River | 243,079 | 0.01% | 99.55% |
| Sri Lanka | Trincomalee | 240,703 | 0.01% | 99.56% |
| Brunei | Tutong | 239,796 | 0.01% | 99.56% |
| Trinidad and Tobago | Sangre Grande | 234,883 | 0.01% | 99.57% |
| Bahamas | West Grand Bahama | 234,844 | 0.01% | 99.57% |
| Palau | Ngardmau | 234,505 | 0.01% | 99.58% |
| South Africa | KwaZulu-Natal | 229,016 | 0.01% | 99.58% |
| Philippines | Isabela | 227,422 | 0.01% | 99.59% |
| Haiti | Sud | 223,024 | 0.01% | 99.59% |
| Palau | Ngatpang | 221,079 | 0.01% | 99.60% |
| Sri Lanka | Mannar | 219,177 | 0.01% | 99.60% |
| Vietnam | Bạc Liêu | 218,106 | 0.01% | 99.61% |
| Vietnam | Hải Phòng | 213,622 | 0.01% | 99.61% |
| Ghana | Western | 212,592 | 0.01% | 99.62% |
| Sri Lanka | Ampara | 211,428 | 0.00% | 99.62% |
| Benin | Mono | 203,914 | 0.00% | 99.63% |
| Puerto Rico | Río Grande | 203,401 | 0.00% | 99.63% |
| Bahamas | South Andros | 194,463 | 0.00% | 99.64% |
| Bahamas | Central Abaco | 192,707 | 0.00% | 99.64% |
| United States | Texas | 189,116 | 0.00% | 99.65% |
| Liberia | Bomi | 188,426 | 0.00% | 99.65% |
| Mexico | Michoacán | 187,838 | 0.00% | 99.66% |
| Indonesia | Bengkulu | 186,931 | 0.00% | 99.66% |
| Thailand | Phatthalung | 185,306 | 0.00% | 99.66% |
| Philippines | Guimaras | 184,410 | 0.00% | 99.67% |
| Liberia | Montserrado | 181,475 | 0.00% | 99.67% |
| China | Hainan | 180,926 | 0.00% | 99.68% |
| Philippines | Aurora | 178,902 | 0.00% | 99.68% |
| Ghana | Greater Accra | 177,456 | 0.00% | 99.69% |
| Liberia | Sinoe | 176,610 | 0.00% | 99.69% |
| Philippines | Pangasinan | 170,535 | 0.00% | 99.69% |
| Puerto Rico | Cabo Rojo | 168,676 | 0.00% | 99.70% |
| United States | Hawaii | 168,203 | 0.00% | 99.70% |
| Liberia | Margibi | 165,551 | 0.00% | 99.71% |
| Palau | Ngchesar | 160,458 | 0.00% | 99.71% |
| Gambia | Banjul | 159,916 | 0.00% | 99.71% |
| Cambodia | Kep | 158,296 | 0.00% | 99.72% |
| Palau | Ngarchelong | 154,804 | 0.00% | 99.72% |
| Philippines | Batangas | 152,292 | 0.00% | 99.72% |
| Vietnam | Tiền Giang | 150,486 | 0.00% | 99.73% |
| Bahamas | Central Andros | 147,383 | 0.00% | 99.73% |
| Jamaica | Trelawny | 146,010 | 0.00% | 99.74% |
| Thailand | Chon Buri | 144,510 | 0.00% | 99.74% |
| Bahamas | Berry Islands | 143,864 | 0.00% | 99.74% |
| Venezuela | Nueva Esparta | 136,589 | 0.00% | 99.75% |
| Mozambique | Maputo City | 135,396 | 0.00% | 99.75% |
| Sri Lanka | Hambantota | 134,930 | 0.00% | 99.75% |
| Haiti | Grand'Anse | 134,352 | 0.00% | 99.76% |
| Timor-Leste | Dili | 132,012 | 0.00% | 99.76% |
| India | Tamil Nadu | 131,375 | 0.00% | 99.76% |
| Trinidad and Tobago | Couva-Tabaquite-Talparo | 131,104 | 0.00% | 99.76% |
| Bahamas | Biminis | 128,870 | 0.00% | 99.77% |
| Trinidad and Tobago | Mayaro/Rio Claro | 128,737 | 0.00% | 99.77% |
| Venezuela | Dependencias Federales | 127,953 | 0.00% | 99.77% |
| Malaysia | Labuan | 127,425 | 0.00% | 99.78% |
| Singapore | North-East | 121,809 | 0.00% | 99.78% |
| Thailand | Narathiwat | 121,215 | 0.00% | 99.78% |
| Dominican Republic | San Pedro de Macorís | 120,743 | 0.00% | 99.79% |
| Palau | Melekeok | 120,190 | 0.00% | 99.79% |
| Cuba | Santiago de Cuba | 117,160 | 0.00% | 99.79% |
| Guinea-Bissau | Bafatá | 116,350 | 0.00% | 99.79% |
| Puerto Rico | Salinas | 116,038 | 0.00% | 99.80% |
| Palau | Ngiwal | 115,965 | 0.00% | 99.80% |
| Puerto Rico | Vega Baja | 115,127 | 0.00% | 99.80% |
| Dominican Republic | María Trinidad Sánchez | 115,002 | 0.00% | 99.80% |
| Bahamas | Moore's Island | 114,844 | 0.00% | 99.81% |
| Cayman Islands | Little Cayman | 114,202 | 0.00% | 99.81% |
| Guyana | Essequibo Islands-West Demerara | 114,109 | 0.00% | 99.81% |
| Trinidad and Tobago | Penal-Debe | 113,884 | 0.00% | 99.82% |
| Thailand | Bangkok Metropolis | 112,084 | 0.00% | 99.82% |
| Philippines | Misamis Oriental | 111,534 | 0.00% | 99.82% |
| Tanzania | Zanzibar North | 111,330 | 0.00% | 99.82% |
| Philippines | Davao del Sur | 110,819 | 0.00% | 99.83% |
| Nicaragua | Managua | 109,219 | 0.00% | 99.83% |
| Martinique | Fort-de-France | 107,380 | 0.00% | 99.83% |
| Jamaica | Hanover | 105,201 | 0.00% | 99.83% |
| Benin | Atlantique | 104,903 | 0.00% | 99.84% |
| Puerto Rico | Naguabo | 104,518 | 0.00% | 99.84% |
| Puerto Rico | Ceiba | 99,651 | 0.00% | 99.84% |
| India | Gujarat | 99,237 | 0.00% | 99.84% |
| Ghana | Central | 98,770 | 0.00% | 99.85% |
| Papua New Guinea | National Capital District | 98,495 | 0.00% | 99.85% |
| Philippines | Biliran | 98,446 | 0.00% | 99.85% |
| Puerto Rico | Carolina | 98,181 | 0.00% | 99.85% |
| Haiti | Ouest | 98,053 | 0.00% | 99.85% |
| Vietnam | Hà Tĩnh | 97,613 | 0.00% | 99.86% |
| United States | Louisiana | 97,068 | 0.00% | 99.86% |
| Colombia | La Guajira | 96,997 | 0.00% | 99.86% |
| Trinidad and Tobago | Chaguanas | 96,093 | 0.00% | 99.86% |
| Venezuela | Trujillo | 90,361 | 0.00% | 99.87% |
| Philippines | Bulacan | 89,458 | 0.00% | 99.87% |
| Bahamas | Inagua | 86,756 | 0.00% | 99.87% |
| Sri Lanka | Kilinochchi | 86,036 | 0.00% | 99.87% |
| Venezuela | Carabobo | 85,849 | 0.00% | 99.87% |
| Puerto Rico | Vieques | 85,117 | 0.00% | 99.88% |
| Puerto Rico | Lajas | 84,044 | 0.00% | 99.88% |
| Suriname | Wanica | 83,053 | 0.00% | 99.88% |
| Cayman Islands | George Town | 81,518 | 0.00% | 99.88% |
| Puerto Rico | Guayama | 81,094 | 0.00% | 99.88% |
| Bonaire, Sint Eustatius and Saba | Bonaire | 80,185 | 0.00% | 99.89% |
| Puerto Rico | Mayagüez | 80,003 | 0.00% | 99.89% |
| Papua New Guinea | Sandaun | 78,742 | 0.00% | 99.89% |
| Vietnam | Bình Định | 78,718 | 0.00% | 99.89% |
| Nicaragua | Rivas | 77,890 | 0.00% | 99.89% |
| Philippines | Bataan | 75,098 | 0.00% | 99.90% |
| Liberia | GrandKru | 74,168 | 0.00% | 99.90% |
| Puerto Rico | Humacao | 73,252 | 0.00% | 99.90% |
| Eritrea | Debubawi Keyih Bahri | 72,613 | 0.00% | 99.90% |
| Bahamas | Hope Town | 71,935 | 0.00% | 99.90% |
| Timor-Leste | Viqueque | 70,035 | 0.00% | 99.90% |
| Puerto Rico | Toa Baja | 69,896 | 0.00% | 99.91% |
| Mexico | Sonora | 68,954 | 0.00% | 99.91% |
| Vanuatu | Shefa | 68,884 | 0.00% | 99.91% |
| El Salvador | La Libertad | 68,083 | 0.00% | 99.91% |
| Thailand | Prachuap Khiri Khan | 67,893 | 0.00% | 99.91% |
| Singapore | North | 67,665 | 0.00% | 99.91% |
| Honduras | Cortés | 66,695 | 0.00% | 99.91% |
| Philippines | Agusan del Norte | 65,376 | 0.00% | 99.92% |
| Antigua and Barbuda | Barbuda | 64,565 | 0.00% | 99.92% |
| New Zealand | Bay of Plenty | 64,067 | 0.00% | 99.92% |
| Singapore | West | 63,070 | 0.00% | 99.92% |
| Palau | Koror | 62,081 | 0.00% | 99.92% |
| Japan | Okinawa | 61,362 | 0.00% | 99.92% |
| Timor-Leste | Ambeno | 61,016 | 0.00% | 99.93% |
| Sri Lanka | Jaffna | 60,668 | 0.00% | 99.93% |
| China | Guangdong | 60,599 | 0.00% | 99.93% |
| Jamaica | Saint Andrew | 59,366 | 0.00% | 99.93% |
| Gambia | Maccarthy Island | 58,808 | 0.00% | 99.93% |
| Cayman Islands | West Bay | 57,837 | 0.00% | 99.93% |
| Bahamas | Mangrove Cay | 57,824 | 0.00% | 99.93% |
| Venezuela | Yaracuy | 57,530 | 0.00% | 99.94% |
| Mayotte | Chirongui | 56,584 | 0.00% | 99.94% |
| Philippines | Pampanga | 55,122 | 0.00% | 99.94% |
| Philippines | Davao del Norte | 53,551 | 0.00% | 99.94% |
| Bahamas | North Abaco | 52,245 | 0.00% | 99.94% |
| Puerto Rico | Fajardo | 50,807 | 0.00% | 99.94% |
| Bahamas | North Andros | 50,794 | 0.00% | 99.94% |
| Timor-Leste | Manatuto | 48,768 | 0.00% | 99.94% |
| Philippines | Compostela Valley | 48,122 | 0.00% | 99.94% |
| Dominican Republic | Azua | 46,996 | 0.00% | 99.95% |
| Puerto Rico | Arecibo | 45,264 | 0.00% | 99.95% |
| Brunei | Belait | 44,735 | 0.00% | 99.95% |
| Trinidad and Tobago | Tobago | 44,557 | 0.00% | 99.95% |
| Bahamas | Cat Island | 44,489 | 0.00% | 99.95% |
| Virgin Islands, U.S. | Saint Croix | 44,264 | 0.00% | 99.95% |
| Costa Rica | Alajuela | 43,995 | 0.00% | 99.95% |
| Mozambique | Gaza | 42,474 | 0.00% | 99.95% |
| Sri Lanka | Gampaha | 41,024 | 0.00% | 99.95% |
| Sri Lanka | Galle | 39,084 | 0.00% | 99.96% |
| Jamaica | Saint Mary | 38,712 | 0.00% | 99.96% |
| Puerto Rico | Guayanilla | 38,240 | 0.00% | 99.96% |
| Puerto Rico | Santa Isabel | 38,117 | 0.00% | 99.96% |
| Togo | Maritime | 37,816 | 0.00% | 99.96% |
| Martinique | Le Trinité | 35,223 | 0.00% | 99.96% |
| Bahamas | South Abaco | 34,183 | 0.00% | 99.96% |
| China | Fujian | 32,424 | 0.00% | 99.96% |
| Philippines | Sarangani | 32,296 | 0.00% | 99.96% |
| Puerto Rico | Ponce | 31,855 | 0.00% | 99.96% |
| Philippines | Ilocos Sur | 31,570 | 0.00% | 99.96% |
| Venezuela | Anzoátegui | 31,376 | 0.00% | 99.96% |
| Guinea-Bissau | Bissau | 31,366 | 0.00% | 99.96% |
| Puerto Rico | San Juan | 31,002 | 0.00% | 99.97% |
| Bahamas | North Eleuthera | 30,570 | 0.00% | 99.97% |
| Bahamas | Long Island | 29,947 | 0.00% | 99.97% |
| Jamaica | Saint James | 29,787 | 0.00% | 99.97% |
| Jamaica | Manchester | 28,755 | 0.00% | 99.97% |
| Nicaragua | Carazo | 28,574 | 0.00% | 99.97% |
| Dominican Republic | Barahona | 28,231 | 0.00% | 99.97% |
| Sri Lanka | Mullaitivu | 28,187 | 0.00% | 99.97% |
| Saint Lucia | Micoud | 26,918 | 0.00% | 99.97% |
| Comoros | Nzwani | 26,472 | 0.00% | 99.97% |
| Philippines | Ilocos Norte | 26,242 | 0.00% | 99.97% |
| Puerto Rico | Patillas | 26,237 | 0.00% | 99.97% |
| India | Odisha | 25,568 | 0.00% | 99.97% |
| Mayotte | Bandraboua | 25,553 | 0.00% | 99.97% |
| Bahamas | East Grand Bahama | 24,825 | 0.00% | 99.97% |
| Guyana | Demerara-Mahaica | 24,785 | 0.00% | 99.98% |
| Mayotte | Mamoudzou | 24,180 | 0.00% | 99.98% |
| Grenada | Carriacou | 23,725 | 0.00% | 99.98% |
| Saint Lucia | Vieux Fort | 23,472 | 0.00% | 99.98% |
| Bahamas | Acklins | 23,097 | 0.00% | 99.98% |
| Haiti | Nord-Ouest | 23,075 | 0.00% | 99.98% |
| Tanzania | Zanzibar West | 23,068 | 0.00% | 99.98% |
| Senegal | Thiès | 22,404 | 0.00% | 99.98% |
| Hong Kong | Yuen Long | 21,981 | 0.00% | 99.98% |
| Timor-Leste | Baucau | 20,626 | 0.00% | 99.98% |
| China | Guangxi | 20,280 | 0.00% | 99.98% |
| Philippines | Siquijor | 20,257 | 0.00% | 99.98% |
| Grenada | Saint Patrick | 20,046 | 0.00% | 99.98% |
| Bahamas | South Eleuthera | 18,336 | 0.00% | 99.98% |
| Bahamas | Central Eleuthera | 18,322 | 0.00% | 99.98% |
| Côte d'Ivoire | Lagunes | 18,240 | 0.00% | 99.98% |
| Virgin Islands, U.S. | Saint Thomas | 17,304 | 0.00% | 99.98% |
| Liberia | Maryland | 17,142 | 0.00% | 99.98% |
| Philippines | La Union | 17,079 | 0.00% | 99.98% |
| Trinidad and Tobago | Point Fortin | 17,064 | 0.00% | 99.98% |
| India | Lakshadweep | 17,056 | 0.00% | 99.99% |
| Puerto Rico | Guánica | 16,551 | 0.00% | 99.99% |
| Puerto Rico | Arroyo | 15,592 | 0.00% | 99.99% |
| Bahamas | New Providence | 15,159 | 0.00% | 99.99% |
| Puerto Rico | Dorado | 15,083 | 0.00% | 99.99% |
| Mayotte | Dembeni | 14,277 | 0.00% | 99.99% |
| Trinidad and Tobago | Port of Spain | 13,840 | 0.00% | 99.99% |
| Indonesia | Jakarta Raya | 13,343 | 0.00% | 99.99% |
| Senegal | Sédhiou | 12,509 | 0.00% | 99.99% |
| Antigua and Barbuda | Saint George | 12,236 | 0.00% | 99.99% |
| Cuba | Ciudad de la Habana | 12,174 | 0.00% | 99.99% |
| Bahamas | Mayaguana | 12,036 | 0.00% | 99.99% |
| Puerto Rico | Peñuelas | 12,023 | 0.00% | 99.99% |
| Puerto Rico | Juana Díaz | 11,974 | 0.00% | 99.99% |
| Vietnam | Thanh Hóa | 11,929 | 0.00% | 99.99% |
| Timor-Leste | Lautém | 11,828 | 0.00% | 99.99% |
| Eritrea | Debub | 11,127 | 0.00% | 99.99% |
| Malaysia | Perlis | 10,956 | 0.00% | 99.99% |
| Mayotte | Koungou | 10,877 | 0.00% | 99.99% |
| Myanmar | Yangon | 10,699 | 0.00% | 99.99% |
| Senegal | Kaolack | 10,512 | 0.00% | 99.99% |
| Saint Lucia | Gros Islet | 10,466 | 0.00% | 99.99% |
| Timor-Leste | Manufahi | 10,058 | 0.00% | 99.99% |
| Cayman Islands | Bodden Town | 9,663 | 0.00% | 99.99% |
| Grenada | Saint Andrew | 9,632 | 0.00% | 99.99% |
| Sri Lanka | Kalutara | 9,344 | 0.00% | 99.99% |
| Bahamas | Exuma | 8,459 | 0.00% | 99.99% |
| United Arab Emirates | Umm Al Qaywayn | 8,313 | 0.00% | 99.99% |
| Saint Vincent and the Grenadines | Grenadines | 8,024 | 0.00% | 99.99% |
| Puerto Rico | Maunabo | 7,762 | 0.00% | 99.99% |
| Djibouti | Obock | 7,650 | 0.00% | 99.99% |
| Senegal | Saint-Louis | 7,537 | 0.00% | 99.99% |
| Jamaica | Portland | 7,412 | 0.00% | 99.99% |
| Grenada | Saint George | 7,287 | 0.00% | 99.99% |
| Trinidad and Tobago | San Fernando | 7,002 | 0.00% | 99.99% |
| British Virgin Islands | Tortola | 6,983 | 0.00% | 99.99% |
| Mayotte | Bandrele | 6,866 | 0.00% | 99.99% |
| Mayotte | Tsingoni | 6,824 | 0.00% | 99.99% |
| Puerto Rico | Culebra | 6,751 | 0.00% | 99.99% |
| Dominican Republic | Peravia | 6,627 | 0.00% | 99.99% |
| British Virgin Islands | Anegada | 6,565 | 0.00% | 99.99% |
| Puerto Rico | Manatí | 6,206 | 0.00% | 99.99% |
| Vietnam | Nghệ An | 6,144 | 0.00% | 100.00% |
| Antigua and Barbuda | Saint Peter | 6,121 | 0.00% | 100.00% |
| Philippines | Metropolitan Manila | 6,027 | 0.00% | 100.00% |
| Saudi Arabia | Makkah | 5,944 | 0.00% | 100.00% |
| Comoros | Njazídja | 5,932 | 0.00% | 100.00% |
| South Africa | Eastern Cape | 5,900 | 0.00% | 100.00% |
| Timor-Leste | Liquiçá | 5,823 | 0.00% | 100.00% |
| Vietnam | Ninh Bình | 5,651 | 0.00% | 100.00% |
| Comoros | Mwali | 5,489 | 0.00% | 100.00% |
| Puerto Rico | Camuy | 5,234 | 0.00% | 100.00% |
| Puerto Rico | Hatillo | 5,096 | 0.00% | 100.00% |
| Somalia | Awdal | 4,704 | 0.00% | 100.00% |
| Saudi Arabia | Jizan | 4,562 | 0.00% | 100.00% |
| Philippines | Cavite | 4,485 | 0.00% | 100.00% |
| Ecuador | Santa Elena | 4,383 | 0.00% | 100.00% |
| Panama | Ngöbe Buglé | 4,362 | 0.00% | 100.00% |
| Puerto Rico | Barceloneta | 4,354 | 0.00% | 100.00% |
| Yemen | Al Hudaydah | 4,098 | 0.00% | 100.00% |
| Grenada | Saint David | 3,947 | 0.00% | 100.00% |
| Turks and Caicos Islands | Providenciales and West Caicos | 3,931 | 0.00% | 100.00% |
| Bonaire, Sint Eustatius and Saba | Saba | 3,726 | 0.00% | 100.00% |
| Puerto Rico | Luquillo | 3,718 | 0.00% | 100.00% |
| Virgin Islands, U.S. | Saint John | 3,706 | 0.00% | 100.00% |
| Mayotte | Kani-Keli | 3,657 | 0.00% | 100.00% |
| Saudi Arabia | Tabuk | 3,648 | 0.00% | 100.00% |
| Antigua and Barbuda | Saint Mary | 3,582 | 0.00% | 100.00% |
| Taiwan | New Taipei | 3,418 | 0.00% | 100.00% |
| Vietnam | Khánh Hòa | 3,316 | 0.00% | 100.00% |
| Nigeria | Ogun | 3,310 | 0.00% | 100.00% |
| Barbados | Christ Church | 3,246 | 0.00% | 100.00% |
| Somalia | Bari | 3,123 | 0.00% | 100.00% |
| Timor-Leste | Bobonaro | 3,064 | 0.00% | 100.00% |
| Bahamas | Crooked Island | 3,036 | 0.00% | 100.00% |
| Antigua and Barbuda | Saint Philip | 2,958 | 0.00% | 100.00% |
| Saint Kitts and Nevis | Saint Thomas Lowland | 2,941 | 0.00% | 100.00% |
| Philippines | South Cotabato | 2,896 | 0.00% | 100.00% |
| Saint Vincent and the Grenadines | Saint George | 2,896 | 0.00% | 100.00% |
| Puerto Rico | Vega Alta | 2,880 | 0.00% | 100.00% |
| Puerto Rico | Aguadilla | 2,656 | 0.00% | 100.00% |
| Venezuela | Aragua | 2,528 | 0.00% | 100.00% |
| Mayotte | M'tsangamouji | 2,527 | 0.00% | 100.00% |
| Mayotte | Sada | 2,461 | 0.00% | 100.00% |
| India | Daman and Diu | 2,372 | 0.00% | 100.00% |
| Puerto Rico | Isabela | 2,211 | 0.00% | 100.00% |
| Puerto Rico | Aguada | 2,180 | 0.00% | 100.00% |
| Vietnam | Bình Thuận | 2,171 | 0.00% | 100.00% |
| Jamaica | Saint Ann | 2,148 | 0.00% | 100.00% |
| Dominican Republic | La Romana | 2,141 | 0.00% | 100.00% |
| Philippines | Camiguin | 2,136 | 0.00% | 100.00% |
| Japan | Kagoshima | 2,124 | 0.00% | 100.00% |
| Bahamas | Harbour Island | 2,000 | 0.00% | 100.00% |
| Morocco | Laâyoune - Boujdour - Sakia El Hamra | 1,924 | 0.00% | 100.00% |
| Yemen | Hajjah | 1,821 | 0.00% | 100.00% |
| Turks and Caicos Islands | North Caicos | 1,790 | 0.00% | 100.00% |
| El Salvador | San Miguel | 1,746 | 0.00% | 100.00% |
| Antigua and Barbuda | Saint John | 1,721 | 0.00% | 100.00% |
| Hong Kong | North | 1,681 | 0.00% | 100.00% |
| United Arab Emirates | Ras Al Khaymah | 1,651 | 0.00% | 100.00% |
| Saint Kitts and Nevis | Saint Peter Basseterre | 1,629 | 0.00% | 100.00% |
| Iran | Hormozgan | 1,510 | 0.00% | 100.00% |
| Puerto Rico | Canóvanas | 1,422 | 0.00% | 100.00% |
| Bahamas | Rum Cay | 1,342 | 0.00% | 100.00% |
| Saint Kitts and Nevis | Saint George Basseterre | 1,215 | 0.00% | 100.00% |
| Philippines | Agusan del Sur | 1,210 | 0.00% | 100.00% |
| Taiwan | Taiwan | 1,149 | 0.00% | 100.00% |
| Hong Kong | Tai Po | 1,119 | 0.00% | 100.00% |
| Singapore | East | 1,081 | 0.00% | 100.00% |
| Turks and Caicos Islands | South Caicos and East Caicos | 1,039 | 0.00% | 100.00% |
| Taiwan | Tainan | 1,032 | 0.00% | 100.00% |
| Mayotte | Boueni | 1,003 | 0.00% | 100.00% |
| Saint Kitts and Nevis | Saint Paul Charlestown | 921 | 0.00% | 100.00% |
| Taiwan | Taipei | 884 | 0.00% | 100.00% |
| Turks and Caicos Islands | Middle Caicos | 847 | 0.00% | 100.00% |
| Hong Kong | Sai Kung | 829 | 0.00% | 100.00% |
| Yemen | Hadramawt | 828 | 0.00% | 100.00% |
| Puerto Rico | Cataño | 822 | 0.00% | 100.00% |
| Puerto Rico | Yabucoa | 761 | 0.00% | 100.00% |
| Saint Lucia | Laborie | 704 | 0.00% | 100.00% |
| Turks and Caicos Islands | Grand Turk | 701 | 0.00% | 100.00% |
| Oman | Muscat | 683 | 0.00% | 100.00% |
| Sudan | Red Sea | 626 | 0.00% | 100.00% |
| United Arab Emirates | Abu Dhabi | 614 | 0.00% | 100.00% |
| Bahamas | Spanish Wells | 589 | 0.00% | 100.00% |
| Djibouti | Djibouti | 548 | 0.00% | 100.00% |
| Saudi Arabia | Al Madinah | 522 | 0.00% | 100.00% |
| Indonesia | Yogyakarta | 514 | 0.00% | 100.00% |
| United Arab Emirates | Fujayrah | 501 | 0.00% | 100.00% |
| Mayotte | Chiconi | 457 | 0.00% | 100.00% |
| Vietnam | Long An | 321 | 0.00% | 100.00% |
| Hong Kong | Islands | 271 | 0.00% | 100.00% |
| Pakistan | Baluchistan | 262 | 0.00% | 100.00% |
| Egypt | Al Bahr al Ahmar | 241 | 0.00% | 100.00% |
| United Arab Emirates | Dubay | 216 | 0.00% | 100.00% |
| Myanmar | Kayin | 205 | 0.00% | 100.00% |
| China | Zhejiang | 161 | 0.00% | 100.00% |
| Senegal | Kolda | 158 | 0.00% | 100.00% |
| Mexico | Baja California | 150 | 0.00% | 100.00% |
| Mayotte | Dzaoudzi | 143 | 0.00% | 100.00% |
| Saudi Arabia | Ash Sharqiyah | 114 | 0.00% | 100.00% |
| Oman | Al Batinah North | 106 | 0.00% | 100.00% |
| Mauritania | Trarza | 102 | 0.00% | 100.00% |
| Bahamas | Black Point | 96 | 0.00% | 100.00% |
| India | Puducherry | 91 | 0.00% | 100.00% |
| Vietnam | Ninh Thuận | 91 | 0.00% | 100.00% |
| Bermuda | Sandys | 88 | 0.00% | 100.00% |
| United Arab Emirates | Ajman | 88 | 0.00% | 100.00% |
| Cayman Islands | Cayman Brac | 87 | 0.00% | 100.00% |
| Oman | Ash Sharqiyah South | 79 | 0.00% | 100.00% |
| Hong Kong | Tuen Mun | 71 | 0.00% | 100.00% |
| Hong Kong | Southern | 69 | 0.00% | 100.00% |
| Guinea-Bissau | Gabú | 52 | 0.00% | 100.00% |
| British Virgin Islands | Virgin Gorda | 50 | 0.00% | 100.00% |
| Oman | Dhofar | 48 | 0.00% | 100.00% |
| Peru | Piura | 44 | 0.00% | 100.00% |
| Bonaire, Sint Eustatius and Saba | Sint Eustatius | 44 | 0.00% | 100.00% |
| Qatar | Al Khor | 44 | 0.00% | 100.00% |
| Taiwan | Kaohsiung | 38 | 0.00% | 100.00% |
| Puerto Rico | Añasco | 35 | 0.00% | 100.00% |
| Puerto Rico | Adjuntas | 33 | 0.00% | 100.00% |
| Bermuda | Hamilton | 17 | 0.00% | 100.00% |
| Bahamas | San Salvador | 7 | 0.00% | 100.00% |
| Hong Kong | Sha Tin | 6 | 0.00% | 100.00% |
| Macao | Macau | 5 | 0.00% | 100.00% |
| Bermuda | Warwick | 5 | 0.00% | 100.00% |
| Bahrain | Southern | 1 | 0.00% | 100.00% |
| Hong Kong | Tsuen Wan | 1 | 0.00% | 100.00% |

Supplemental Figure 1. Global Mangrove Carbon Stocks by Year.

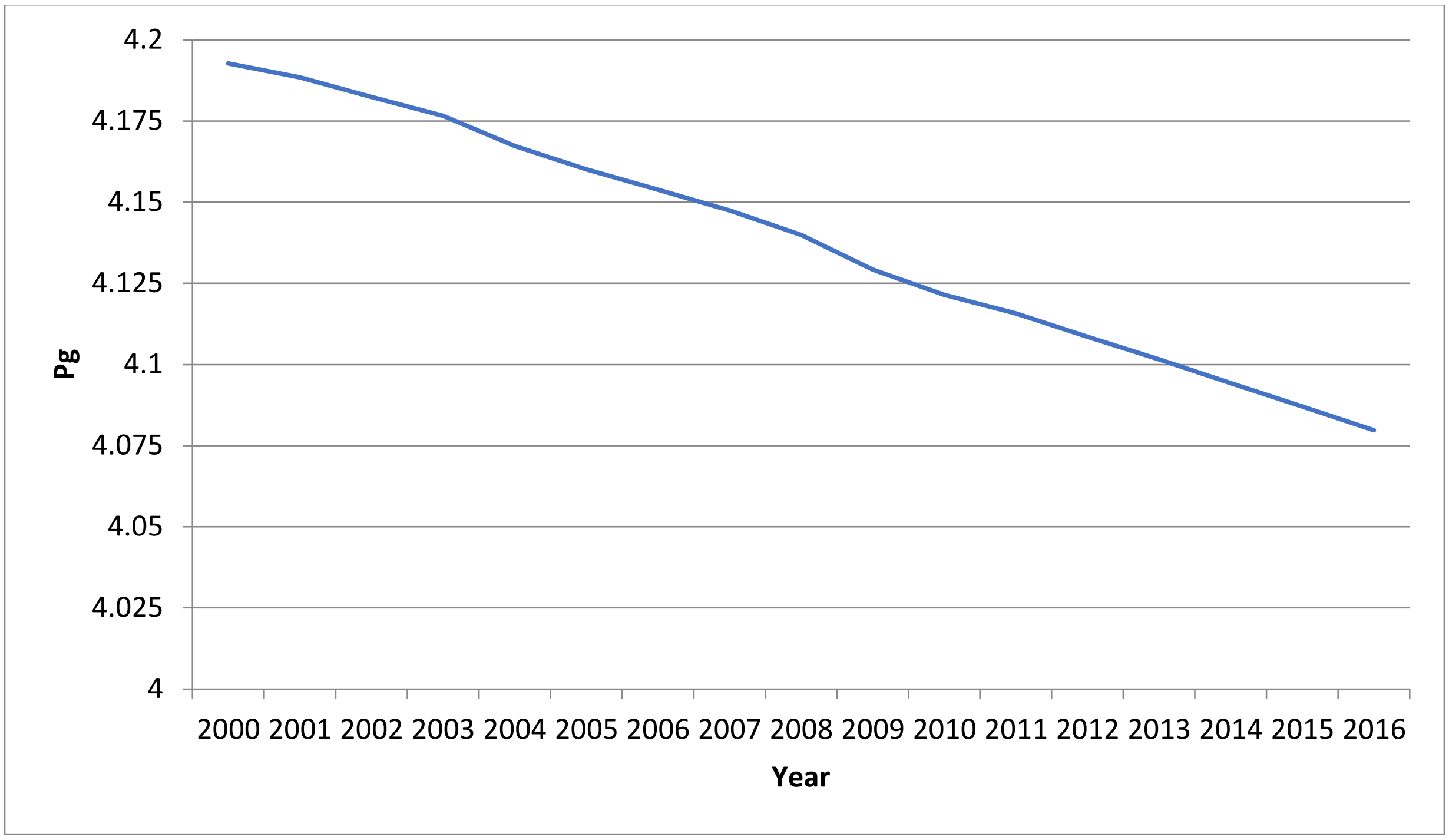

EQ 5 adjusted for mean of all equations by less than 1% and used for 2000 to 2012 global estimate and future predictions for 2013 to 2016. The predictions use the following linear model were $y = -0.0074x + 4.2901$, $R^2$ 0.99, were x is the last two years of the year and is the global mangrove carbon stock.